\documentclass[sigconf]{acmart}

\setcopyright{none}
\renewcommand\footnotetextcopyrightpermission[1]{}

\usepackage{tikz}
\usetikzlibrary{arrows.meta,positioning}
\usepackage{algorithm}
\usepackage{algpseudocode}
\usepackage{subcaption}
\usepackage{booktabs}
\usepackage{enumitem}
\usepackage{array}
\usepackage{graphicx}
\usepackage{tabularx}
\usepackage{float}

\begin{document}

\title{Verifiable, Privacy-Preserving and Collaborative Multi-Objective Optimization with NSGA-II on Blockchain}
\author{Goshgar Can Ismayilov}
\affiliation{%
  \city{Istanbul}
  \country{Turkey}}
\email{goshgar.ismayilov@boun.edu.tr}

\begin{abstract}
Collaborative multi-objective optimization allows independent problem solvers to contribute their solutions to a shared problem. However, their validity cannot be reliably verified in open settings if the parties are reluctant to disclose their solutions. Therefore, this paper proposes a novel zk-NSGA-II protocol for verifiable, privacy-preserving and collaborative multi-objective optimization on blockchain. To the best of our knowledge, this is the first work that executes Non-Dominated Sorting Genetic Algorithm II (NSGA-II) on the RISC Zero Zero-Knowledge Virtual Machine (zkVM) where evolutionary generations are certified by succinct proofs that are verified on-chain. Problem solvers maintain their own independent populations as the set of solutions in the decision space and their corresponding objective vectors in the objective space. They contribute to a global on-chain Pareto front by aggregating their local off-chain Pareto fronts according to the rule of non-domination which admits a contribution if it is non-dominated with respect to the existing contributions in the global front. To prevent fabricated contributions, population states across generations are linked through hash chains. The correctness of the zk-NSGA-II protocol is first evaluated on twelve problems from the ZDT and DTLZ benchmarks. Second, its feasibility is shown in a real-world portfolio optimization case study with the Hang Seng dataset. Performance on these problems is measured with nine different metrics across three complementary perspectives for optimization, zero-knowledge proof and blockchain. The empirical findings indicate that trustless collaborative multi-objective optimization is feasible with the existing zero-knowledge and blockchain technologies.
\end{abstract}


\maketitle
\pagestyle{plain}

\section{Introduction}
\label{sec:introduction}

Consider a financial investor who asks several asset managers to construct portfolios, balancing the expected return against the risk. No portfolio is superior in both respects in principle since higher returns generally come at the cost of higher risks. Therefore, the investor is interested in the efficient frontier jointly formed from the portfolios of all managers. Although the asset managers are willing to report the returns and risks of their portfolios to form this frontier, they are reluctant to reveal the exact asset weights to the public, which are part of their hidden proprietary strategies. On the other hand, the investor needs to verify these reports since a portfolio with unusually high return at low risk might stem from a sound strategy or merely a clever presentation of a mediocre one. In this scenario, the investor can at least be assured that the portfolio is indeed the product of a mutually-agreed set of operations without observing the weights. This is the issue of privacy-preserving but collaborative multi-objective optimization (MOO) among mutually-distrusting parties, with the conflicting objectives of return maximization and risk minimization. This work addresses privacy with zero-knowledge proof (ZKP), public verifiability and trust with blockchain, MOO with NSGA-II \cite{nsga2} and collaboration with frontier aggregation.

The first key question about the scenario above is \textit{who will form the frontier with individual portfolios}, which naturally raises the trust issue. The simplest answer is the investor itself. However, the managers would have to trust the investor about not maliciously changing contributions, not favoring certain ones or not colluding with the other managers. Relying on a neutral coordinator would also shift the trust issue somewhere else, rather than solving it. Our work incorporates blockchain to properly address this issue. Blockchain is a distributed ledger that is shared across all nodes in the network \cite{nakamoto2008}. Smart contracts extend the idea of distributed ledger with public, self-executing and deterministic programs with states and functions \cite{buterin2014}. For our scenario, blockchain offers several notable advantages. First, every contribution is permanently recorded and attributed to its contributor for auditability. Second, the rules of frontier aggregation are publicly-encoded and agreed. Third, participation is open to anyone at any time without the approval of an authority. In the literature, there exist works that focus on optimization and blockchain at the same time to replace hash puzzles with Proof-of-Useful-Work  (PoUW) \cite{shibata2019, fitzi2022} and to execute genetic algorithms \cite{bizzaro2020}. However, these works reveal contributions for validation, which the managers in our scenario would not accept. 

The second key question is \textit{how to convince without revealing weights}. In this work, we incorporate ZKP, which is a special cryptographic technique that allows a prover to convince a verifier about a certain statement \cite{goldwasser1989}. Modern proof systems make this practical for general-purpose computations. For instance, zkSNARKs produce succinct and non-interactive proofs that can be verified in milliseconds off-chain \cite{bensasson2014} while zkSTARKs remove the need for a trusted setup \cite{bensasson2018}. In addition, zkVMs such as RISC Zero \cite{bruestle2023} prove the execution of ordinary programs and thus eliminate the need to design hand-crafted arithmetic circuits. ZKPs offer several critical advantages. First and most importantly, what should remain private remains private. Second, due to the soundness of proof systems, valid proofs cannot be forged for incorrect results. Third, proofs are succinct and cheap to verify on-chain regardless of how long the actual optimization is. These properties have been exploited for verifiable machine learning applications \cite{liu2021, chen2024, sun2024}. However, to the best of our knowledge, integration with multi-objective evolutionary algorithms (MOEAs) has not been considered.

\subsection{Contributions}
\label{sec:contributions}

This paper makes the following contributions to the literature:

\begin{itemize}
    \item \textbf{zk-NSGA-II as verifiable, privacy-preserving and collaborative NSGA-II in zkVM.} As the core contribution of this work, we propose zk-NSGA-II which executes NSGA-II on the RISC Zero zkVM off-chain and certifies the result with succinct proofs to verify on-chain. In zk-NSGA-II, problem owners deploy their problem instances as smart contracts while problem solvers maintain independent and asynchronous populations to solve these instances collaboratively. 
    
    \item \textbf{Aggregation of local fronts into a global Pareto front.} We formulate the privacy-preserving MOO from the federated perspective so that the mutually-distrusting solvers maintain their own local Pareto fronts off-chain and contribute to a single global Pareto front on-chain. The contract admits a new solution if it is valid and no solution on the global front dominates it. It also removes the solutions that the new one dominates from the global front. During front aggregation, populations are never exposed to the public except the objective vector of the non-dominated solution submitted to the contract.  

    \item \textbf{Extensive experimental study.} We evaluate zk-NSGA-II on twelve synthetic problems from the ZDT and DTLZ benchmarks to justify its correctness. In the experiments, we use performance measures from three complementary perspectives: (i) number of non-dominated solutions, Schott's spacing, inverted generational distance and hypervolume ratio for MOO, (ii) zkVM cycles, proving time, verification time and proof size for ZKP and (iii) function-level gas consumption for the blockchain perspective. 

    \item \textbf{A privacy-preserving portfolio optimization case study.} We apply zk-NSGA-II to the Hang Seng portfolio optimization to demonstrate its performance on a real-world problem. In the case, the solvers report the risk and return of their portfolios while the portfolio weights remain private. Our findings on both synthetic and real problems justify the feasibility of zk-NSGA-II while further improvements are necessary for practicality such as reduction in proving time.
\end{itemize}

\subsection{Paper Organization}
\label{sec:organization}

The remainder of the work is organized in the following way. Section \ref{sec:background} provides the necessary background on MOO, ZKP and smart contracts. Section \ref{sec:related-work} reviews related work on privacy-preserving evolutionary computation, verifiable computation on blockchain and decentralized optimization on blockchain. Section \ref{sec:zknsga2} proposes the key contributions of the work by introducing the zk-NSGA-II protocol on a zkVM. Section \ref{sec:setup} describes the experimental setup including the benchmark problems, real-world case study and performance measures. Section \ref{sec:results} reports the empirical results to show correctness and efficiency of zk-NSGA-II on numerous scenarios. Finally, Section \ref{sec:conclusion} concludes the work. 

\section{Background}
\label{sec:background}

The proposed protocol lies at the intersection of three domains as blockchain, ZKP and MOO. Therefore, we briefly provide background for the concepts of Pareto-domination, zkVMs and on-chain proof verification in this section. We also establish the terminology and common notations used throughout the work, which are listed in Tables \ref{tab:abbreviations} and \ref{tab:notation}, respectively.

\begin{table}[t]
\caption{Common abbreviations used throughout the paper, listed in alphabetical order.}
\label{tab:abbreviations}
\small
\begin{tabularx}{\columnwidth}{@{}lX@{}}
\toprule
\textbf{Abbreviation} & \textbf{Meaning} \\
\midrule
DTLZ & Deb--Thiele--Laumanns--Zitzler test suite \\
EVM & Ethereum virtual machine \\
HVR & Hypervolume ratio \\
IGD & Inverted generational distance \\
MOEA & Multi-objective evolutionary algorithm \\
MOO & Multi-objective optimization \\
NS & Number of non-dominated solutions \\
NSGA-II & Non-dominated sorting genetic algorithm II \\
RISC-V & Reduced instruction set computer, fifth generation \\
SHA-256 & Secure hash algorithm, 256-bit digest \\
SNARK & Succinct non-interactive argument of knowledge \\
SS & Schott's spacing \\
STARK & Scalable transparent argument of knowledge \\
ZDT & Zitzler--Deb--Thiele test suite \\
ZKP & Zero-knowledge proof \\
zkVM & Zero-knowledge virtual machine \\
\bottomrule
\end{tabularx}
\end{table}

\begin{table}[t]
\caption{Common mathematical notation used throughout the paper, grouped by MOO, collaborative optimization, blockchain and portfolio optimization.}
\label{tab:notation}
\small
\begin{tabularx}{\columnwidth}{@{}lX@{}}
\toprule
\textbf{Symbol} & \textbf{Description} \\
\midrule
\multicolumn{2}{@{}l}{\emph{Multi-objective optimization}} \\
$\mathbf{x}$, $n$ & Decision vector $(x_1, \dots, x_n)$ and number of decision variables \\
$\Omega$ & Decision space $\prod_{j=1}^{n} [l_j, u_j]$ defined by variable bounds \\
$\mathbf{F}(\mathbf{x})$ & Objective vector $(f_1(\mathbf{x}), f_2(\mathbf{x}))$ \\
$\mathbf{a} \prec \mathbf{b}$ & Objective vector $\mathbf{a}$ Pareto-dominates $\mathbf{b}$ \\
$\mathbf{a} \preceq \mathbf{b}$ & Objective vector $\mathbf{a}$ weakly dominates $\mathbf{b}$ \\
$\mathcal{P}^{\ast}$ & True Pareto front \\
$N$, $G$, $t$ & Population size, number of generations, generation index \\
$P_t$, $Q_t$ & Parent and offspring populations at generation $t$ \\
$p_c$, $p_m$ & Crossover and mutation probabilities \\
$\mathbf{r}$ & Reference point for the hypervolume measure \\
\midrule
\multicolumn{2}{@{}l}{\emph{Collaborative optimization}} \\
$S$, $i$ & Number of solvers, solver index \\
$a_i$ & Blockchain address of solver $i$ \\
$\mathbf{x}^{\ast}_{t}$ & Solution selected from the local front \\
$L_i^t$ & Local front (off-chain) of solver $i$ at generation $t$ \\
$\mathcal{G}$ & Global (on-chain) front \\
$K$ & Maximum size of the global front \\
\midrule
\multicolumn{2}{@{}l}{\emph{Zero-knowledge proof and blockchain}} \\
$\Gamma$ & NSGA-II guest program proven in the zkVM \\
$\pi_t$ & Groth16 proof of the generation from $P_t$ to $P_{t+1}$ \\
$o_t$ & Public outputs $(a_i, h_t, h_{t+1}, \mathbf{F}(\mathbf{x}^{\ast}_{t+1}))$ of $\pi_t$ \\
$H(\cdot)$ & SHA-256 hash function \\
$h_t$ & State hash of the population $P_t$ at generation $t$ \\
$\mathrm{head}[a_i]$ & Latest state hash of solver $i$ stored on-chain \\
\midrule
\multicolumn{2}{@{}l}{\emph{Portfolio optimization}} \\
$\mathbf{w}$ & Portfolio weights, $w_j \geq 0$, $\sum_j w_j = 1$ \\
$\boldsymbol{\mu}$, $\boldsymbol{\Sigma}$ & Mean asset returns and covariance matrix \\
$\mathbf{w}^{\top}\boldsymbol{\Sigma}\mathbf{w}$ & Portfolio risk (variance) \\
$\boldsymbol{\mu}^{\top}\mathbf{w}$ & Portfolio expected return \\
\bottomrule
\end{tabularx}
\end{table}

\subsection{Multi-Objective Optimization}
\label{sec:bg-moo}

MOO concerns a group of problems in which multiple conflicting objectives need to be optimized simultaneously \cite{deb2001}. Without loss of generality, a minimization problem with $m$ objectives can be formulated in the following way:
\begin{equation}
\label{eq:moo}
\begin{aligned}
\min_{\mathbf{x} \in \Omega} \quad & \mathbf{F}(\mathbf{x}) = \bigl(f_1(\mathbf{x}), f_2(\mathbf{x}), \dots, f_m(\mathbf{x})\bigr) \\
\text{subject to} \quad & c_j(\mathbf{x}) \leq 0, \quad j = 1, \dots, C \\
& e_l(\mathbf{x}) = 0, \quad l = 1, \dots, E
\end{aligned}
\end{equation}
where $\mathbf{x} = (x_1, \dots, x_n)$ is the decision vector, $\Omega$ is the decision space while $c_j$ and $e_l$ are the inequality and equality constraints, respectively. In MOO with conflicting objectives, there is generally no single solution that minimizes all these objectives at once. Instead, the solutions are ranked through the principle of Pareto domination on their objective vectors. An objective vector $\mathbf{a} \in \mathbb{R}^m$ \textit{weakly-dominates} another objective vector $\mathbf{b} \in \mathbb{R}^m$ ($\mathbf{a} \preceq \mathbf{b}$) if $a_i \leq b_i$ for all objectives $i \in \{1, \dots, m\}$ and \textit{Pareto-dominates} it ($\mathbf{a} \prec \mathbf{b}$) if, in addition, $a_i < b_i$ for at least one objective $i$. A solution is referred to as Pareto-optimal if no other solution dominates it. The set of all such Pareto-optimal solutions in the objective space forms the Pareto front $\mathcal{P}$. The goal of MOO is therefore to find a set of non-dominated solutions that approximates $\mathcal{P}$ as closely as possible. In this work, we consider only $m = 2$ objectives.

NSGA-II \cite{nsga2} is among the most widely used MOEAs \cite{ismayilov2018, ismayilov2020}. NSGA-II maintains a parent population $P_t$ with the size of $N$ and produces an offspring population $Q_t$ with the same size in every generation $t$. The parent and offspring populations are then merged into $R_t = P_t \cup Q_t$ of size $2N$ and ranked through the fast non-dominated sorting into successive fronts $\mathcal{F}_1, \mathcal{F}_2, \dots$, where $\mathcal{F}_1$ refers to the non-dominated solutions. The next population $P_{t+1}$ is filled front by front until all $N$ solutions are selected. For the comparison of solutions in the same front, the crowding distance is used, which prefers the solutions in less crowded regions of the objective space. Overall, NSGA-II is an elitist approach since it always preserves the non-dominated solutions against the dominated ones. 

\subsection{Zero-Knowledge Proof}
\label{sec:bg-zk}

Introduced by Goldwasser, Micali and Rackoff \cite{goldwasser1989}, a ZKP is a protocol in which a prover convinces a verifier about the correctness of a statement. A ZKP satisfies three properties: \emph{completeness} guarantees that a prover with a valid witness always convinces a verifier; \emph{soundness} guarantees that no dishonest prover with an invalid witness can convince the verifier; and \emph{zero-knowledge} ensures that the verifier eventually learns nothing beyond the validity of the statement. Modern proof systems are non-interactive where proof generation and verification can be separately carried out without the need for interaction between the prover and the verifier. Out of these proof systems, zkSTARKs \cite{bensasson2018} require no trusted setup with the trade-off of larger proof size while zkSNARKs \cite{bensasson2014} require a trusted setup but produce constant-size proofs of a few hundred bytes, which are cheap to verify on a blockchain.

General-purpose proof systems express the computation as an arithmetic circuit, which is known as arithmetization. Designing a dedicated circuit for every computation is error-prone, especially for complex algorithms such as NSGA-II that involve loops, data-dependent branching, sorting and floating-point arithmetic. A zkVM addresses this issue by proving the correct execution of an arbitrary program, written in a programming language. In this work, we adopt the RISC Zero zkVM \cite{bruestle2023} that executes programs compiled to the RISC-V instruction set. The proven program (i.e., \emph{guest}) is written in Rust and a host supplies the guest with private inputs to run it. In zk-NSGA-II, the guest executes one NSGA-II generation on a private population while the hash chain for the previous and new populations is maintained on-chain for integrity.

\subsection{Smart Contracts}
\label{sec:bg-contracts}

In its purest form, a blockchain is an append-only database that is shared across all the nodes in a peer-to-peer network \cite{nakamoto2008}. In permissionless blockchains such as Bitcoin and Ethereum, anyone can join the network with pseudonymous addresses, read the ledger and submit transactions. Blockchain groups these separate transactions into blocks and chains those blocks with the hash of their predecessor block. This chaining enables transaction history to be tamper-proof since even a minimal change in any past transaction propagates into all subsequent blocks. The nodes in the network agree on blocks through a consensus protocol which selects the node to create the next block. However, the integrity, availability and transparency that a blockchain provides come at a price. First, all data on a public blockchain is visible to every node, which poses risks to privacy \cite{ismayilov2025ptts}. Second, replication and re-execution of transactions by every node make computation and storage on-chain more expensive than off-chain, which motivates moving heavy and privacy-sensitive computations out of chain.

Smart contracts are self-executing programs that can be deployed on a blockchain \cite{buterin2014}. A contract has a state and a set of functions that nodes can call through transactions. Since every node executes the same contract functions on the same state, the results of these calls are deterministic and publicly verifiable. Ethereum is a contract-supporting blockchain where contracts are written in Solidity, compiled to bytecode and executed by the Ethereum Virtual Machine (EVM). Each operation in the EVM is linked to a fixed amount of gas that the calling node needs to pay. The total gas cost of a function is the sum of the gas costs of all operations it executes. From the gas cost viewpoint, running a heavy computation directly on-chain is impractical while verifying a succinct proof of that computation is still affordable. In this work, our contracts verify the proofs of problem solvers to ensure the correct computation of NSGA-II while the populations still remain off-chain.

\section{Related Work}
\label{sec:related-work}

In this section, we review the literature of privacy-preserving evolutionary computation, verifiable computation on blockchain and decentralized optimization on blockchain. To the best of our knowledge, no prior work combines ZKP-based correct execution of a MOEA with the trustless front aggregation of mutually-distrusting solvers while still preserving the privacy of their decision variables.

\subsection{Privacy-Preserving Evolutionary Computation}
\label{sec:rw-evolutionary}

Privacy-preserving evolutionary computation is addressed even before the blockchain era through secure multi-party computation where several parties jointly run an evolutionary algorithm. The work \cite{sakuma2007} proposes a genetic algorithm for distributed combinatorial optimization, especially for the traveling salesman problem. The following work \cite{funke2010} extends the secure computation to the entire MOEA. The work \cite{xu2021} addresses the federated data-driven evolutionary optimization when the data is distributed across separate servers. It is later extended to multi-objective and many-objective problems as well \cite{xu2022}. However, secure computation requires all parties to be available during the entire optimization and assumes that they honestly follow the protocol rules. In addition, none of these approaches allows verifying that the algorithm has been faithfully executed. ZKP specifically addresses this gap by certifying the correctness of a computation. The work \cite{angel2022} proposes Otti, which is a general-purpose compiler for zkSNARKs that supports numerical optimization problems. It provides efficient arithmetization, especially for linear programming, semi-definite programming and stochastic gradient descent instances. In contrast to these works, zk-NSGA-II enables solvers to run NSGA-II independently on their own machines in a privacy-preserving and verifiable manner.

\subsection{Verifiable Computation on Blockchain}
\label{sec:rw-verifiable}

Verifiable computation allows verifier nodes on blockchain to check the result of a computation that is outsourced to a prover node. In the literature, there currently exist several systems that make this practical with succinct proofs in terms of proof size \cite{groth2016} and with no trusted setup \cite{bensasson2018}. For an optimistic approach, the work \cite{teutsch2023} moves computation off-chain by only resolving disputes on-chain by introducing challengers to disagree. For zero-knowledge approaches, the work \cite{kosba2016} relies on privacy-preserving smart contracts while another work \cite{eberhardt2018} provides a toolbox to compile off-chain computations into proofs that are later verified inside contracts. For multi-chain systems, the work \cite{xie2022} also verifies the consensus of one chain on another with zkBridge. The works for zkCNN \cite{liu2021}, ZKML \cite{chen2024} and zkLLM \cite{sun2024} also make the verification of large machine and deep learning computations practical. However, note that all these systems express the computation in terms of dedicated arithmetic circuits. zkVMs remove this requirement by proving executions of entire programs. For instance, the RISC Zero zkVM \cite{bruestle2023} allows custom Rust programs to be efficiently verified on Ethereum. However, no prior work proves the faithful execution of a MOEA, which is the gap that zk-NSGA-II addresses.

\subsection{Decentralized Optimization on Blockchain}
\label{sec:rw-distributed}

In the literature, there currently exist several works that couple blockchain with optimization. The work \cite{shibata2019} proposes Proof-of-Search where the node submitting the best candidate solution to an optimization problem is rewarded as the next block miner. Another work \cite{bizzaro2020} extends this idea to Proof-of-Evolution so that the miners run genetic algorithms and cooperate by sharing their resulting solutions. A similar scheme is also proposed for real-life combinatorial optimization problems as PoUW \cite{todorovic2022}. The work \cite{fitzi2022} proposes the first provably secure PoUW protocol that relies on a novel local search algorithm, called Doubly Parallel Local Search. The work \cite{cao2024} notes that most works focus on specific problems or algorithms and presents a generic PoUW framework for arbitrary problem formulations with a security analysis. Beyond the consensus mechanisms, smart contracts are used to coordinate decentralized optimization without a trusted coordinator in domains such as energy management \cite{yang2021}. However, none of these approaches addresses the issue of solution privacy or collaborative MOO for front aggregation. 

More recently, ZKPs are introduced to protect the privacy of learning and optimization systems. The following work \cite{wang2025} proposes zkFL which proves the correctness of gradient aggregation in federated learning where no blockchain nodes learn the local models. Another work \cite{ismayilov2025} also proposes to couple evolutionary computation with programmable cryptography on blockchain that preserves the privacy of populations. Our protocol, zk-NSGA-II, builds on this line of research but differs in three critical aspects. First, it specifically targets MOO and executes a complete NSGA-II generation inside a general-purpose zkVM. Second, it links consecutive proofs of solvers through hash chains for integrity. Third, as model aggregation in federated learning, it enables the aggregation of objective spaces of mutually-distrusting solvers into a global front on-chain while still keeping their decision variables hidden. 

\section{zk-NSGA-II: Verifiable NSGA-II in zkVM}  
\label{sec:zknsga2}

This section presents zk-NSGA-II, which runs NSGA-II inside the RISC Zero zkVM off-chain and verifies the resulting executions on-chain. The main motivation of the protocol is to move the entire evolutionary computation off-chain and enable blockchain nodes to verify succinct proofs of that computation. zk-NSGA-II has four core properties: \textit{verifiable} since every solution in the global front is certified to be the result of a faithful NSGA-II execution, \textit{privacy-preserving} since the populations and hence the decision variables never leave the local solver machines, \textit{collaborative} since mutually-distrusting solvers first evolve their own populations independently and later contribute to a global front with no trusted coordinator and finally \textit{multi-objective} since the guest program executes a complete NSGA-II generation with the non-dominated sorting and crowding distance operators. The pseudocode of the solver, the guest program and the contract is given in Appendix \ref{app:pseudocode}.

\subsection{Actors}
\label{sec:actors}
The system involves four types of actors: problem owner who poses the optimization problem, solver who searches for solutions, prover who certifies the computation of solver and on-chain contract to verify computation and aggregate multiple local fronts to a single global front. For this work, we assume that solvers are also provers. 

\subsubsection{Problem Owner} The problem owners seek high-quality solutions for their multi-objective problems with objective functions and constraints. They are responsible for deploying contracts which bind problem instances to the maximum size of the global front. Once deployed, the contracts are ready to orchestrate optimization. 

\subsubsection{Solvers} Solvers are mutually-distrusting participants that run NSGA-II by maintaining their own populations and therefore their own local Pareto fronts off-chain. The decision variables in populations never leave the solvers, but proofs reporting the correctness of computation are submitted to the contract, which binds the objective values of a locally non-dominated solution to the solver address. Solvers advance one generation per proof and initial submissions might start from an arbitrary population. This enables solvers to explore different regions of the search space for collaborative optimization.

\subsubsection{Provers} Provers execute the NSGA-II program on zkVM with solver inputs and generate proofs to certify the correctness of computation. Here, two different designs are possible where the first design lets solvers delegate proving to third-party provers if the solvers lack necessary hardware. However, provers necessarily observe solver inputs or might refuse to provide service. The second design assumes solvers as provers, so sensitive inputs remain on their own machines. The soundness of the proof system used in the work prevents any prover from certifying an incorrect computation. 

\subsubsection{On-Chain Contract}
The contract is responsible for verifying solver submissions and admitting their solutions to the global front. It maintains two key states: a single global on-chain front as the aggregation of multiple local off-chain fronts and hash chains of solvers to ensure that the current populations are always generated from the previous ones. It admits a new solution if the current solutions at the global front cannot dominate that new solution. Once admitted, the solutions that are dominated by that new solution are removed.

\subsection{Trust Assumptions}
\label{sec:trust}

The security of zk-NSGA-II assumes dishonest problem solvers with no trusted party. The additional assumptions are listed as:

\begin{enumerate}[leftmargin=*]
\item \textbf{Soundness and zero-knowledge of the proof system.} It is computationally infeasible to produce a valid proof for an execution that differs from the actual program. In addition, a proof reveals nothing about the private inputs of the program beyond its public outputs, which keeps the populations of the solvers private. However, the Groth16 system \cite{groth2016} needs a circuit-specific trusted setup. We assume that at least one participant of that setup is honest and destroys secret randomness. 

\item \textbf{Collision and preimage resistance of the hash function.} It is also computationally infeasible to find two different inputs with the same SHA-256 output. This assumption must hold for the state hash chain of each solver. In addition, it is computationally infeasible to recover a population from its public state hash. Since the state hash is not salted, this relies on the high entropy of the population, which consists of $N \cdot n$ real-valued decision variables. 

\item \textbf{Correctness of the NSGA-II program.} The program implements NSGA-II by correctly specifying objective functions and constraints. A proof can only certify that this program is executed faithfully, but not that the program itself is correct.

\item \textbf{Correct and live execution of the blockchain.} The underlying blockchain platform is expected to operate reliably by executing the smart contracts exactly as written and eventually including valid transactions into its blocks. 

\end{enumerate}

\noindent
Since solvers act as their own provers (Section \ref{sec:actors}), no assumption about third-party provers is required. If proving were delegated to independent prover nodes, they would have to be trusted with the confidentiality of the solver inputs. 

\subsection{Architectural Design}
\label{sec:architecture}

\begin{figure*}[t]
\centering
\def\st#1{\guillemotleft#1\guillemotright}
\begin{tikzpicture}[
  font=\scriptsize, >={Stealth[length=2mm]},
  line width=0.5pt,
  comp/.style={draw=black, fill=white, align=center, inner sep=3pt, minimum height=0.75cm},
  cls/.style={draw=black, fill=white, text width=5.5cm, align=left, inner sep=3pt},
  call/.style={->},
  ret/.style={->, densely dashed},
  stepnum/.style={circle, draw=black, fill=black, text=white, font=\tiny\bfseries, inner sep=0pt, minimum size=3.6mm},
  lbl/.style={font=\scriptsize, inner sep=2pt}
]
\newcommand{\umlnode}[4]{%
  \fill[black!12] (#1,#4) -- ++(0.2,0.2) -- (#3+0.2,#4+0.2) -- (#3,#4) -- cycle;
  \fill[black!20] (#3,#4) -- (#3+0.2,#4+0.2) -- (#3+0.2,#2+0.2) -- (#3,#2) -- cycle;
  \fill[black!3] (#1,#2) rectangle (#3,#4);
  \draw (#1,#2) rectangle (#3,#4);
  \draw (#1,#4) -- ++(0.2,0.2) -- (#3+0.2,#4+0.2) -- (#3,#4);
  \draw (#3+0.2,#4+0.2) -- (#3+0.2,#2+0.2) -- (#3,#2);}
\newcommand{\compicon}[1]{%
  \begin{scope}[shift={(#1.north east)}, line width=0.4pt]
    \draw[fill=white] (-0.36,-0.1) rectangle (-0.12,-0.34);
    \draw[fill=white] (-0.42,-0.15) rectangle (-0.3,-0.2);
    \draw[fill=white] (-0.42,-0.24) rectangle (-0.3,-0.29);
  \end{scope}}
\newcommand{\lockicon}[1]{%
  \begin{scope}[shift={(#1)}, xshift=-0.12cm, line width=0.45pt]
    \draw (-0.055,0.02) -- (-0.055,0.07) arc (180:0:0.055) -- (0.055,0.02);
    \draw[fill=black!70] (-0.09,-0.1) rectangle (0.09,0.03);
  \end{scope}}
\newcommand{\stepR}[3]{
  \node[stepnum] (sn) at (#1) {#2};
  \node[lbl, anchor=west, align=left] at (sn.east) {#3};}
\newcommand{\stepL}[3]{
  \node[stepnum] (sn) at (#1) {#2};
  \node[lbl, anchor=east, align=right] at (sn.west) {#3};}
\newcommand{\stepA}[3]{
  \node[stepnum] (sn) at (#1) {#2};
  \node[lbl, anchor=south] at (sn.north) {#3};}

\begin{scope}[shift={(0.3,0.3)}]\umlnode{0}{-6.2}{8.5}{0}\end{scope}
\begin{scope}[shift={(0.15,0.15)}]\umlnode{0}{-6.2}{8.5}{0}\end{scope}
\umlnode{0}{-6.2}{8.5}{0}
\node[anchor=north west, align=left] at (0.1,-0.08)
  {\textbf{Solver machine $i$} ($i = 1, \dots, S$)};
\node[anchor=north east] (priv) at (8.4,-0.08)
  {\textit{private: populations}};
\lockicon{priv.west}

\node[comp, text width=7.4cm] (H) at (4.25,-1.2)
  {\st{application}\ \textbf{Host}};

\node[comp, text width=2.3cm] (DB) at (1.45,-4.6)
  {\st{datastore}\\ \textbf{Population Store}};

\draw[fill=black!8] (3.6,-5.9) rectangle (8.2,-2.4);
\node[anchor=south east] at (8.15,-5.85) {\textbf{RISC Zero zkVM}};

\node[comp, text width=1.8cm] (G) at (4.9,-3.4)
  {\st{guest program}\\ \textbf{NSGA-II}};
\node[comp, text width=1.8cm] (PR) at (4.9,-5.0)
  {\st{process}\\ \textbf{Prover}};

\umlnode{11.15}{-6.2}{17.25}{0}
\node[anchor=north west] at (11.25,-0.08) {\textbf{Ethereum}};
\node[anchor=north east] at (17.15,-0.08) {\textit{public: $\pi_t$, $o_t$, $\mathcal{G}$}};

\node[cls, align=center, anchor=north] (PF) at (14.2,-0.7)
  {\st{contract}\ \textbf{Pareto Front}};
\node[cls, anchor=north] (PFa) at (PF.south)
  {$K$: capacity \\
   $\mathrm{head}[a_i]$: latest population hash of solver $i$\\
   $\mathcal{G}$: global Pareto front};
\node[cls, anchor=north] (PFo) at (PFa.south)
  {\texttt{bool} $\leftarrow$ \texttt{submit}($\pi_t$, $o_t$)\\
   $\mathcal{G}$ $\leftarrow$ \texttt{getFront}()};

\node[cls, align=center, anchor=north] (V) at (14.2,-4.0)
  {\st{contract}\ \textbf{Groth16 Verifier}};
\node[cls, anchor=north] (Vo) at (V.south)
  {\texttt{valid} / \texttt{invalid} $\leftarrow$ \texttt{verify}($\pi_t, o_t$)};

\node[anchor=south] at (14.2,1.65) {\st{actor}\ \textbf{Problem Owner}};

\draw[call] (13.4,1.6) -- (13.4,0 |- PF.north);
\stepL{13.4,0.9}{1}{deploy the ParetoFront and Groth16Verifier contracts}

\draw[<->] (0.9,0 |- DB.north) -- (0.9,0 |- H.south);
\stepR{0.9,-2.85}{2}{load the current\\population $P_{t}$}
\stepR{0.9,-3.75}{7}{store the new\\population $P_{t+1}$}

\draw[call] (4.3,0 |- H.south) -- (4.3,0 |- G.north);
\stepL{4.3,-1.95}{3}{execute NSGA-II on $P_{t}$}

\draw[call] (4.9,0 |- G.south) -- (4.9,0 |- PR.north);
\stepR{4.9,-4.22}{4}{prove the execution trace}

\draw[ret] (5.5,0 |- G.north) -- (5.5,0 |- H.south);
\stepR{5.5,-2.0}{5}{return the new\\population $P_{t+1}$}

\draw[ret] (PR.east) -| (7.9,0 |- H.south);
\stepL{7.9,-3.4}{6}{return the proof $\pi_t$\\and outputs $o_t$}

\draw[call] (H.east) -- (H.east -| PF.west);
\stepA{10.0,-1.2}{8}{submit $(\pi_t, o_t)$}
\node[lbl, anchor=north, font=\scriptsize\itshape] at (10.0,-1.4) {public data only};

\draw[call] (13.3,0 |- PFo.south) -- (13.3,0 |- V.north);
\draw[ret] (15.2,0 |- V.north) -- (15.2,0 |- PFo.south);
\stepL{13.3,-3.4}{9}{verify the proof $\pi_t$}
\node[lbl, anchor=west, align=left] at (15.2,-3.45) {return \texttt{valid} / \texttt{invalid}};

\draw[ret] (15.0,0 |- PF.north) -- (15.0,1.6);
\stepR{15.0,0.9}{10}{read the global front $\mathcal{G}$}
\end{tikzpicture}
\caption{Architectural design of zk-NSGA-II with off-chain solver machines and on-chain Ethereum contracts. The problem owner first deploys the Pareto front and verifier contracts. Off-chain, the host application of each solver loads its private population $P_{t}$ from the population store and executes the NSGA-II guest program inside the RISC Zero zkVM. The prover then generates a proof $\pi_t$ of this execution and the host stores the new population $P_{t+1}$. Only the proof and the public outputs $o_t = (a_i, h_t, h_{t+1}, \mathbf{F}(\mathbf{x}^{\ast}_{t+1}))$ are submitted on-chain while the populations stay hidden. On-chain, the Pareto Front contract delegates the proof verification to the Groth16 verifier contract, checks the correctness of the hash chain and updates the global front $\mathcal{G}$. The problem owner can read the global front at any time. Steps 2--9 are repeated by every solver in every generation.}
\label{fig:architecture}
\end{figure*}
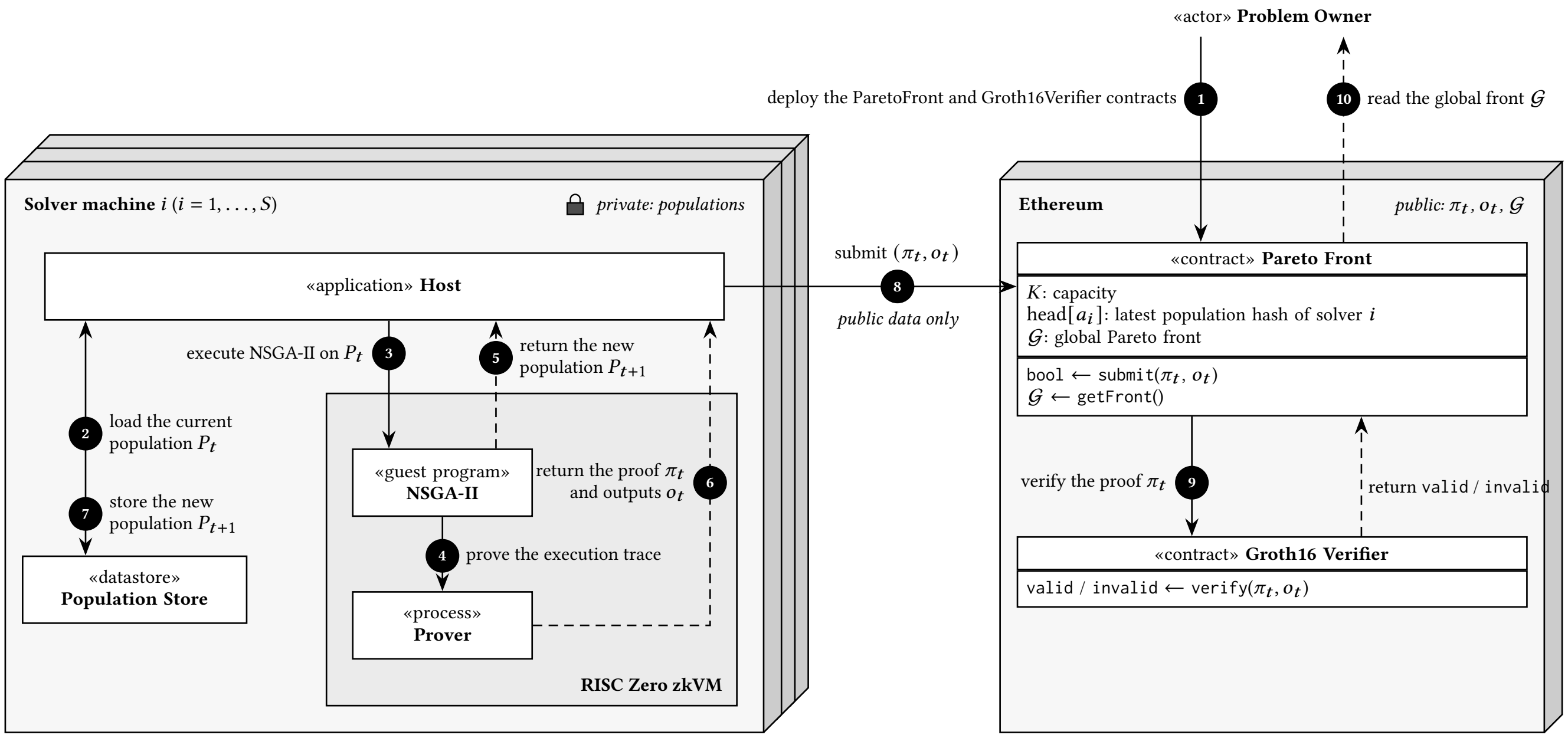

The architectural design of zk-NSGA-II separates computation and verification into two different layers, as shown in Figure \ref{fig:architecture}. Off-chain, three components work interdependently where the host application is untrusted and orchestrates the optimization-centric computation while the population store is responsible for maintaining the latest version of the population across generations and the RISC Zero zkVM executes the NSGA-II guest program and proves its execution. On-chain, the problem owner deploys two contracts for a single problem instance. The Pareto Front contract holds the state such as the capacity $K$ of the global front, the latest population hash $\mathrm{head}[a_i]$ of solvers and the global front $\mathcal{G}$. The Groth16 verifier is a stateless contract that only checks proofs.

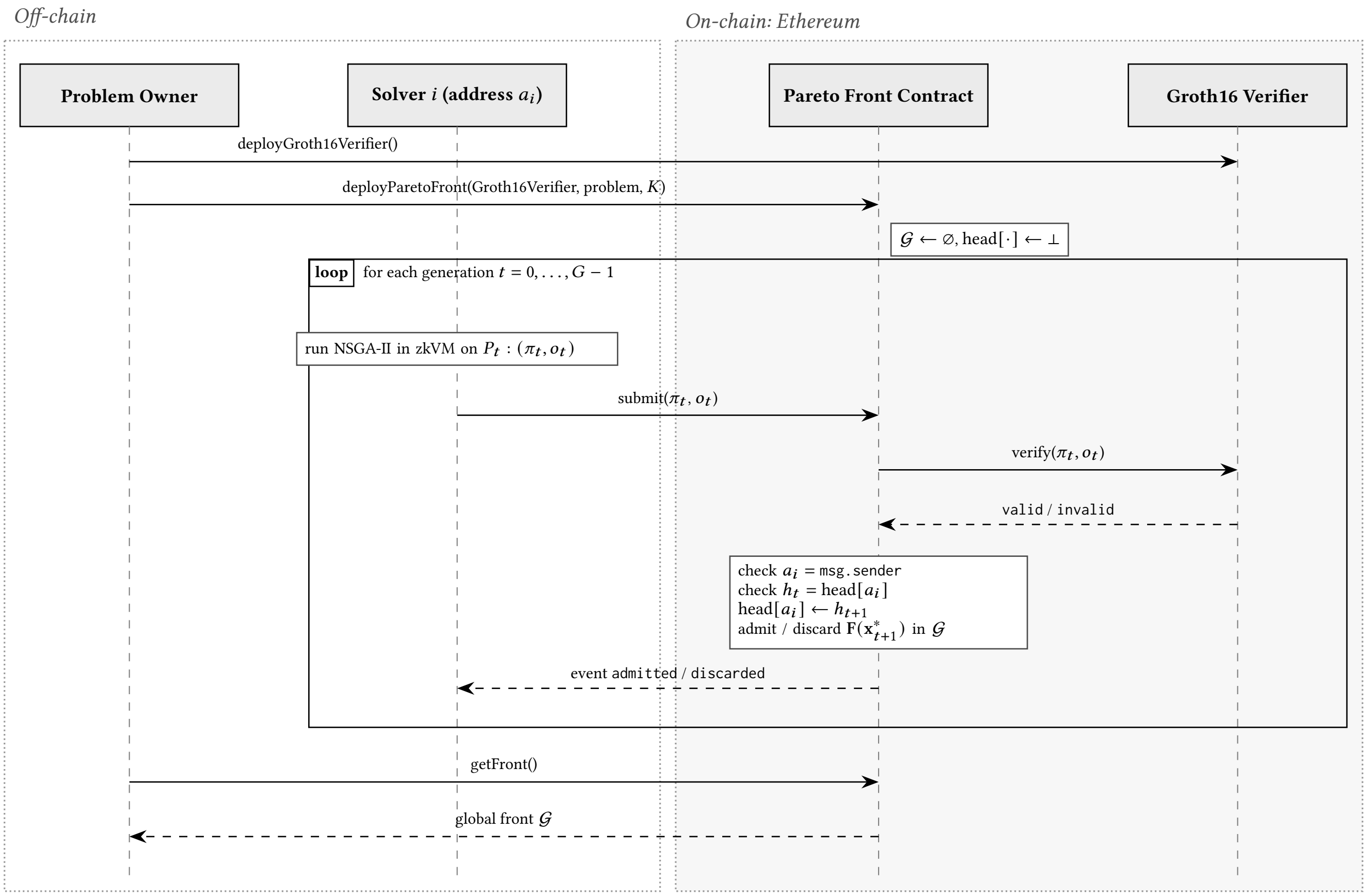
\begin{figure*}[t]
\centering
\begin{tikzpicture}[
  font=\small, >={Stealth[length=2.2mm]},
  line width=0.5pt,
  actor/.style={draw=black, fill=black!8, minimum width=2.8cm, minimum height=0.8cm, align=center, font=\footnotesize\bfseries},
  life/.style={dashed, black!50, line width=0.4pt},
  msg/.style={->, black},
  ret/.style={->, dashed, black},
  lbl/.style={font=\scriptsize, above, align=center},
  selfbox/.style={draw=black!70, fill=white, font=\scriptsize, align=left, inner sep=3pt},
  region/.style={draw=black!40, densely dotted, line width=0.6pt},
  regiontitle/.style={font=\small\itshape, text=black!70, anchor=south west}
]

\def\xO{0}     
\def\xS{4.2}   
\def\xP{9.6}   
\def\xV{14.2}  
\def\yEnd{-10.0}

\draw[region] (-1.6,0.7) rectangle (6.8,\yEnd-0.2);
\node[regiontitle] at (-1.6,0.75) {Off-chain};
\fill[black!3] (7.0,0.7) rectangle (15.8,\yEnd-0.2);
\draw[region] (7.0,0.7) rectangle (15.8,\yEnd-0.2);
\node[regiontitle] at (7.0,0.75) {On-chain: Ethereum};

\node[actor] (O) at (\xO,0) {Problem Owner};
\node[actor] (S) at (\xS,0) {Solver $i$ (address $a_i$)};
\node[actor] (P) at (\xP,0) {Pareto Front Contract};
\node[actor] (V) at (\xV,0) {Groth16 Verifier};
\foreach \x in {\xO,\xS,\xP,\xV} {
  \draw[life] (\x,-0.4) -- (\x,\yEnd);
}

\draw[msg] (\xO,-0.85) -- node[lbl, pos=0.17] {deployGroth16Verifier()} (\xV,-0.85);
\draw[msg] (\xO,-1.4) -- node[lbl] {deployParetoFront(Groth16Verifier, problem, $K$)} (\xP,-1.4);
\node[selfbox, anchor=west] at (\xP+0.15,-1.85) {$\mathcal{G} \gets \varnothing$, $\mathrm{head}[\cdot] \gets \bot$};

\draw[black] (2.3,-2.1) rectangle (15.6,-8.1);
\node[draw=black, fill=white, font=\scriptsize\bfseries, anchor=north west, inner sep=2pt] (L) at (2.3,-2.1) {loop};
\node[font=\scriptsize, anchor=west] at (L.east) {for each generation $t = 0, \dots, G-1$};

\node[selfbox, text width=3.9cm] at (\xS,-3.25) {%
  run NSGA-II in zkVM on $P_t: (\pi_t, o_t)$};

\draw[msg] (\xS,-4.1) -- node[lbl] {submit($\pi_t$, $o_t$)} (\xP,-4.1);
\draw[msg] (\xP,-4.8) -- node[lbl] {verify($\pi_t, o_t$)} (\xV,-4.8);
\draw[ret] (\xV,-5.5) -- node[lbl] {\texttt{valid} / \texttt{invalid}} (\xP,-5.5);

\node[selfbox, text width=3.6cm] at (\xP,-6.5) {%
  check $a_i = \texttt{msg.sender}$\\
  check $h_t = \mathrm{head}[a_i]$\\
  $\mathrm{head}[a_i] \gets h_{t+1}$\\
  admit / discard $\mathbf{F}(\mathbf{x}^{\ast}_{t+1})$ in $\mathcal{G}$};

\draw[ret] (\xP,-7.6) -- node[lbl] {event \texttt{admitted} / \texttt{discarded}} (\xS,-7.6);

\draw[msg] (\xO,-8.8) -- node[lbl] {getFront()} (\xP,-8.8);
\draw[ret] (\xP,-9.5) -- node[lbl] {global front $\mathcal{G}$} (\xO,-9.5);
\end{tikzpicture}
\caption{Sequence diagram of zk-NSGA-II. The problem owner first deploys the Groth16 verifier contract and then the Pareto Front contract for a problem instance. In every generation, a solver runs NSGA-II on its private population in zkVM and submits the resulting proof $\pi_t$ together with the public outputs $o_t = (a_i, h_t, h_{t+1}, \mathbf{F}(\mathbf{x}^{\ast}_{t+1}))$. The Pareto Front contract delegates the proof verification to the Groth16 verifier contract and later checks the solver and the hash chain. It admits or discards the objective vector. The problem owner can read the global front $\mathcal{G}$ afterwards.}
\label{fig:sequence}
\end{figure*}

In Figure \ref{fig:architecture}, the problem owner first deploys the contracts to initiate an optimization session (step 1). For every generation, the host loads the current population from the population store (step 2) and executes the guest program on it (step 3). The prover then generates a corresponding Groth16 proof $\pi_t$ (step 4). The zkVM returns the new population to the host (step 5) together with the proof and the public outputs (step 6). The host stores the new population for the next generation in the population store (step 7). The host then submits only $(\pi_t, o_t)$ to the Pareto Front contract (step 8). The contract delegates the proof verification operation to the Groth16 Verifier contract (step 9). If the proof is valid, the Pareto Front contract performs the additional operations for orchestration. The problem owner later reads the global front (step 10). The problem solvers available for this problem instance repeat steps 2--9 independently. Figure \ref{fig:sequence} shows the same interaction as a sequence diagram from contract deployment to reading the global front.

\subsection{Off-Chain Proof Generation}
\label{sec:proof-generation}

In zk-NSGA-II, each solver runs NSGA-II off-chain as the guest program inside the RISC Zero zkVM (steps 2--7 in Figure \ref{fig:architecture}). For the first proof, the solver can start from an arbitrary initial population $P_0$ of $N$ solutions. For every subsequent proof, it continues from the current population $P_t$. In both cases, the guest receives the population as a private input, checks that all solutions lie within the variable bounds, re-evaluates their objective vectors and then evolves the population to the next generation. In each generation, parents are chosen by binary tournament selection \cite{goldberg1991}. They are recombined with simulated binary crossover \cite{deb1995} and mutated by polynomial mutation \cite{deb1996} until $N$ offspring are produced. The parents and offspring are then merged into a population of size $2N$ and sorted into fronts with fast non-dominated sorting. The first front of $P_{t+1}$ forms the local front $L_i^{t+1}$ of the solver, from which the objective vector of a single solution $\mathbf{x}^{\ast}_{t+1}$ is selected to be reported. Only the objective vector $\mathbf{F}(\mathbf{x}^{\ast}_{t+1})$ and the hashes of $P_t$ and $P_{t+1}$ are submitted on-chain while the populations with the decision variables remain hidden. The execution is proven by RISC Zero and compressed into a constant-size Groth16 proof $\pi_t$ for on-chain verification as:
\begin{equation}
\label{eq:prove}
(\pi_t, o_t, P_{t+1}) \leftarrow \mathrm{Prove}\bigl(\Gamma; P_t, a_i\bigr)
\end{equation}
where $\Gamma$ is the NSGA-II guest program and $o_t = (a_i, h_t, h_{t+1}, \mathbf{F}(\mathbf{x}^{\ast}_{t+1}))$ are the public outputs. The proof $\pi_t$ shows the correctness of $o_t$ while revealing nothing about $P_t$. 

\subsection{On-Chain Proof Verification}
\label{sec:proof-verification}

After a solver generates a Groth16 proof for its generation off-chain, the proof is submitted to the Pareto Front contract for on-chain verification (steps 8--9 in Figure \ref{fig:architecture}). The contract does not implement verification itself but rather delegates it to the Groth16 verifier contract of the RISC Zero zkVM \cite{bruestle2023}. Note that a valid proof only certifies the faithfulness of off-chain NSGA-II execution. Therefore, the contract needs to perform two additional post-verification checks. First, the correctness of the hash chain must be verified (i.e., the previous state hash must be equal to the current head $\mathrm{head}[a_i]$ stored on-chain). Second, the blockchain address of the problem solver (i.e., $\texttt{msg.sender}$) must be verified. If these checks successfully pass, the head of the solver is updated accordingly. Formally, a submission is verified if and only if:
\begin{equation}
\label{eq:accept}
\begin{aligned}
& \mathrm{Verify}\bigl(\Gamma; \pi_t, o_t\bigr) \\
& \;\wedge\; a_i = \texttt{msg.sender} \\
& \;\wedge\; \bigl(\mathrm{head}[a_i] = \bot \;\vee\; h_t = \mathrm{head}[a_i]\bigr)
\end{aligned}
\end{equation}
where $\bot$ is the empty head of a solver. The objective vector submitted together with the proof is compared against the global front with respect to the non-domination principle. The contract admits the new solution if it is non-dominated and removes the existing solutions that the new solution dominates.

\subsection{Hash Chain for Computation Integrity}
\label{sec:hash-chain}       

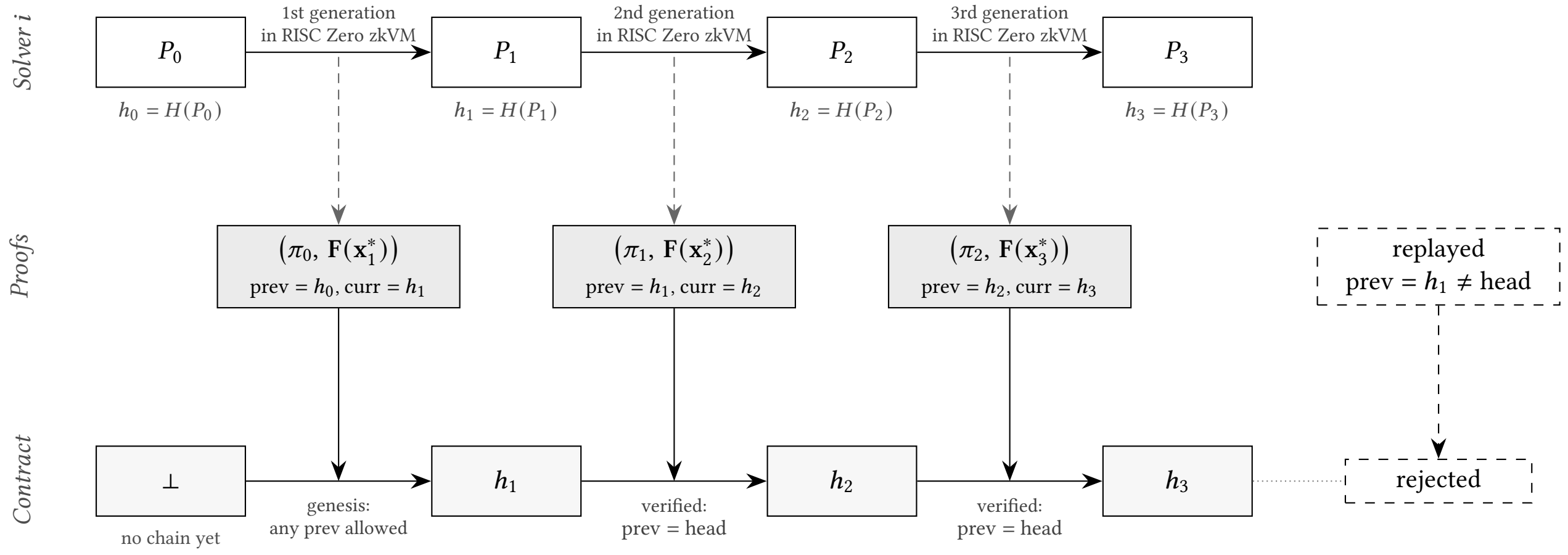
\begin{figure*}[t]
\centering
\begin{tikzpicture}[
  font=\small, >={Stealth[length=2.2mm]},
  state/.style={draw=black, minimum width=1.6cm, minimum height=0.75cm, fill=white},
  proof/.style={draw=black, align=center, fill=black!8, inner sep=3pt, minimum width=2.6cm},
  head/.style={draw=black, minimum width=1.6cm, minimum height=0.75cm, fill=black!3},
  bad/.style={draw=black, dashed, align=center, fill=white, inner sep=3pt, minimum width=2.6cm},
  hash/.style={font=\scriptsize, text=black!70},
  steplabel/.style={font=\scriptsize, text=black!75, align=center},
  lane/.style={font=\small\itshape, text=black!70, rotate=90, anchor=center}
]
\node[lane] at (-1.6, 0)    {Solver $i$};
\node[lane] at (-1.6, -2.3) {Proofs};
\node[lane] at (-1.6, -4.6) {Contract};

\node[state] (P0) at (0, 0)    {$P_0$};
\node[state] (P1) at (3.6, 0)  {$P_1$};
\node[state] (P2) at (7.2, 0)  {$P_2$};
\node[state] (P3) at (10.8, 0) {$P_3$};
\node[hash, below=1pt of P0] {$h_0 = H(P_0)$};
\node[hash, below=1pt of P1] {$h_1 = H(P_1)$};
\node[hash, below=1pt of P2] {$h_2 = H(P_2)$};
\node[hash, below=1pt of P3] {$h_3 = H(P_3)$};
\draw[->] (P0) -- node[steplabel, above] {1st generation\\in RISC Zero zkVM} (P1);
\draw[->] (P1) -- node[steplabel, above] {2nd generation\\in RISC Zero zkVM} (P2);
\draw[->] (P2) -- node[steplabel, above] {3rd generation\\in RISC Zero zkVM} (P3);

\foreach \idx/\cx/\pv in {1/1.8/0, 2/5.4/1, 3/9.0/2} {
  \node[proof] (J\idx) at (\cx, -2.3) {$\left(\pi_{\pv},\, \mathbf{F}(\mathbf{x}^{\ast}_{\idx})\right)$\\[1pt] {\scriptsize $\mathrm{prev} = h_{\pv}$, $\mathrm{curr} = h_{\idx}$}};
  \draw[->, densely dashed, black!60] (\cx, -0.05) -- (J\idx);
}

\node[head] (H0) at (0, -4.6)    {$\bot$};
\node[head] (H1) at (3.6, -4.6)  {$h_1$};
\node[head] (H2) at (7.2, -4.6)  {$h_2$};
\node[head] (H3) at (10.8, -4.6) {$h_3$};
\node[hash, below=1pt of H0] {no chain yet};
\node[hash, below=1pt of H1, align=center] {};
\node[hash, below=1pt of H2, align=center] {};
\node[hash, below=1pt of H3, align=center] {};
\draw[->] (H0) -- node[steplabel, below=2pt] {genesis:\\any prev allowed} (H1);
\draw[->] (H1) -- node[steplabel, below=2pt] {verified:\\$\mathrm{prev} = \mathrm{head}$} (H2);
\draw[->] (H2) -- node[steplabel, below=2pt] {verified:\\$\mathrm{prev} = \mathrm{head}$} (H3);
\draw[->] (J1) -- (1.8, -4.6);
\draw[->] (J2) -- (5.4, -4.6);
\draw[->] (J3) -- (9.0, -4.6);

\node[bad] (Jx) at (13.6, -2.3) {replayed\\$\mathrm{prev} = h_1 \neq \mathrm{head}$};
\node[bad, minimum width=2.0cm] (R) at (13.6, -4.6) {rejected};
\draw[->, dashed, black] (Jx) -- (R);
\draw[densely dotted, black!50] (H3) -- (R);
\end{tikzpicture}
\caption{Hash chain linking the population of a solver $i$ with address $a_i$ across three generations. Off-chain, the solver advances its population by a single generation and submits its proof $\pi_t$, together with the objective vector $\mathbf{F}(\mathbf{x}^{\ast}_{t+1})$ of the selected non-dominated solution from its local front. On-chain, the contract maintains the latest state hash for every solver. The first submission starts its chain from an arbitrary state, but every subsequent submission is verified only if its previous state is correct. The state is replaced with the new hash. A mismatched proof is simply rejected.}
\label{fig:hash-chain}
\end{figure*}

While a generation proof certifies that the population $P_{t+1}$ is correctly derived from the population $P_t$, it says nothing about where $P_t$ comes from. An adversarial solver could easily start every proof from a population of its own choice without following the generational approach of NSGA-II. In this case, each proof in isolation would be valid while the sequence of consecutive proofs does not correspond to any actual run of NSGA-II. The most direct solution would be to store populations on-chain. However, this would reveal the decision variables to the public and also incur a storage cost that grows with the population size and number of decision variables. To address this issue, zk-NSGA-II links the proofs of each solver with a chain of hashes, as illustrated in Figure \ref{fig:hash-chain}.

The hash of a population is defined in the following way:
\begin{equation}
\label{eq:state-hash}
h_t = \mathrm{H}\bigl(t \,\|\, s_t \,\|\, N \,\|\, n \,\|\, \mathbf{x}_1 \,\|\, \cdots \,\|\, \mathbf{x}_N\bigr)
\end{equation}
where $\|$ denotes concatenation, $s_t$ is the state of the pseudo-random number generator and $\mathbf{x}_1, \dots, \mathbf{x}_N$ are the decision vectors of $P_t$. The objective vectors are deliberately excluded since the guest program can re-compute from the decision variables. In each proof, the guest program receives $P_{t}$ as a private input, computes $h_{t}$, evolves the population by a single generation and computes the hash of the resulting population. On-chain, the contract stores only the latest population hash $\mathrm{head}[a_i]$ of every solver $i$. After verifying a proof, the contract considers the submission verified only if its previous state hash equals $\mathrm{head}[a_i]$ and then replaces $\mathrm{head}[a_i]$ with the new state hash. A caveat here is that the first submission of a solver can start from an arbitrary state without enforcing an initial starting population. In this way, solvers can explore solutions in distinct regions, improving the diversity of decentralized optimization. However, from then on, every proof must continue exactly where the previous proof ended. The hash chain provides several guarantees for our work. First, a chain clearly corresponds to one uninterrupted NSGA-II run from its genesis population to the final population. Second, it prevents proofs from being replayed or forked since the proofs are rejected when their previous state hash does not match. Third, the privacy of populations is preserved while computations carried out are still verifiable. 

\subsection{From Local Fronts to Global Front}         \label{sec:local-global}  

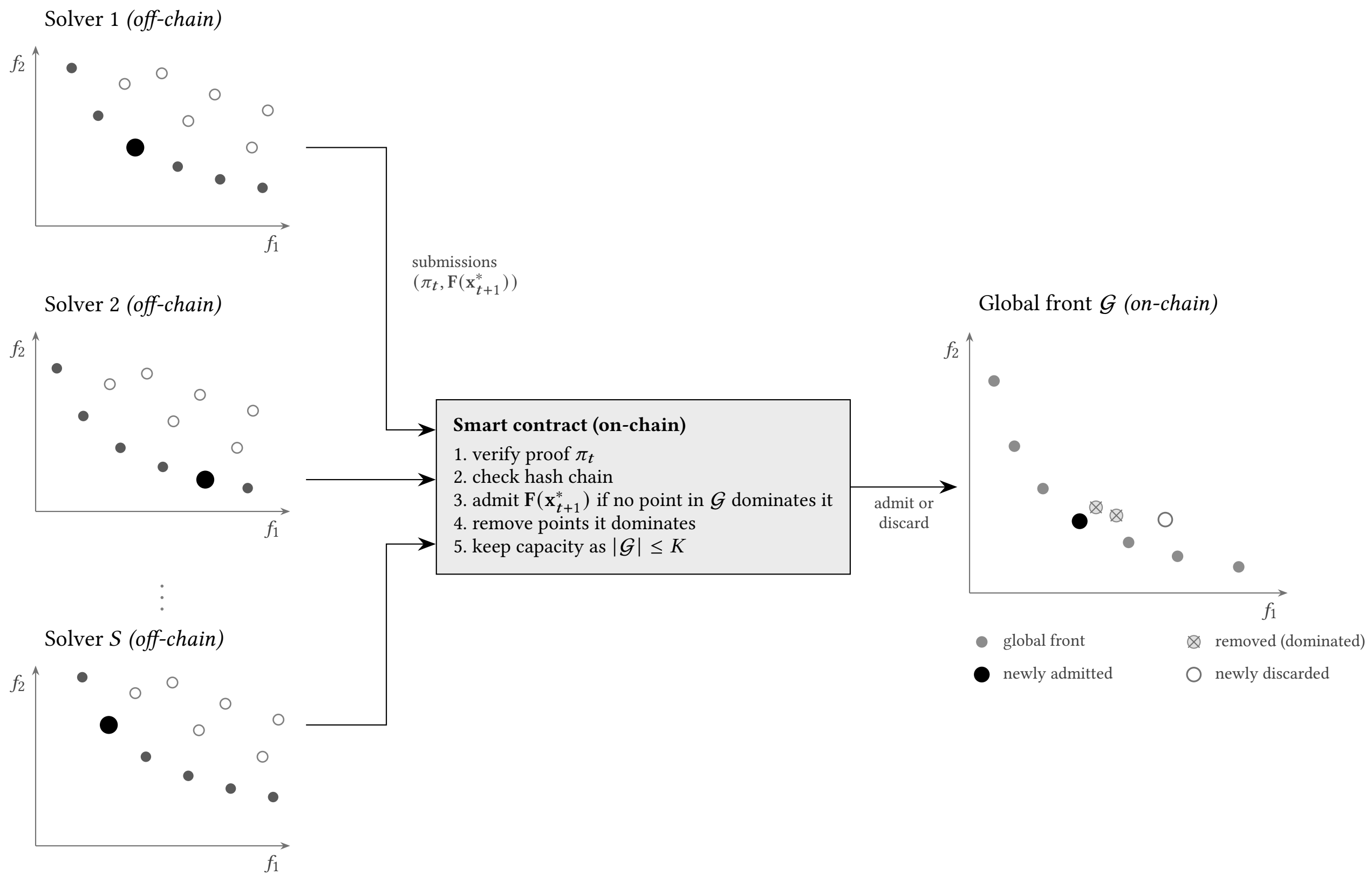
\begin{figure*}[t]
\centering
\begin{tikzpicture}[
  font=\small, >={Stealth[length=2.2mm]},
  panelaxis/.style={->, >={Stealth[length=1.2mm]}, black!55},
  localpt/.style={fill=black!65},
  localdom/.style={draw=black!50, fill=white, line width=0.5pt},
  reportpt/.style={fill=black, draw=white, line width=0.5pt},
  globalpt/.style={fill=black!45},
  removedpt/.style={draw=black!45, fill=black!12},
  discardpt/.style={draw=black!55, fill=white, line width=0.6pt},
  contract/.style={draw=black, fill=black!8, align=left, inner sep=6pt},
  note/.style={font=\scriptsize, text=black!75},
  paneltitle/.style={font=\small, anchor=south west}
]
\foreach \k/\name/\sy/\off/\rx/\ry in {%
    1/{Solver 1}/3.5/0.14/0.8/0.6,
    2/{Solver 2}/0.0/0.00/1.6/0.3,
    3/{Solver $S$}/-4.1/0.24/0.45/0.9} {
  \begin{scope}[shift={(0,\sy)}, scale=1.3]
    \draw[panelaxis] (0,0) -- (2.4,0) node[note, below left] {$f_1$};
    \draw[panelaxis] (0,0) -- (0,1.7) node[note, below left] {$f_2$};
    \foreach \px/\py in {0.2/1.35, 0.45/0.9, 0.8/0.6, 1.2/0.42, 1.6/0.3, 2.0/0.22} {
      \fill[localpt] (\px+\off, \py+\off) circle (1.4pt);
    }
    \foreach \px/\py in {0.7/1.2, 1.05/1.3, 1.3/0.85, 1.55/1.1, 1.9/0.6, 2.05/0.95} {
      \draw[localdom] (\px+\off, \py+\off) circle (1.4pt);
    }
    \filldraw[reportpt] (\rx+\off, \ry+\off) circle (2.6pt);
    \coordinate (port\k) at (2.55, \ry+\off);
    \node[paneltitle] at (0, 1.75) {\name\ \textit{(off-chain)}};
  \end{scope}
}
\node[font=\large, text=black!50] at (1.55, -0.95) {$\vdots$};

\node[contract, font=\footnotesize] (C) at (7.45, 0.3) {%
  \textbf{Smart contract (on-chain)}\\[2pt]
  1.\ verify proof $\pi_t$\\
  2.\ check hash chain\\
  3.\ admit $\mathbf{F}(\mathbf{x}^{\ast}_{t+1})$ if no point in $\mathcal{G}$ dominates it\\
  4.\ remove points it dominates\\
  5.\ keep capacity as $|\mathcal{G}| \le K$};

\draw[->] (port1) -- (4.3, 0 |- port1) |- ([yshift=7mm]C.west);
\draw[->] (port2) -- (C.west |- port2);
\draw[->] (port3) -- (4.3, 0 |- port3) |- ([yshift=-7mm]C.west);
\node[note, anchor=west, align=left] at (4.5, 2.9) {submissions\\$(\pi_t, \mathbf{F}(\mathbf{x}^{\ast}_{t+1}))$};

\begin{scope}[shift={(11.45,-1.0)}]
  \draw[panelaxis] (0,0) -- (3.9,0) node[note, below left] {$f_1$};
  \draw[panelaxis] (0,0) -- (0,3.2) node[note, below left] {$f_2$};
  \fill[removedpt] (1.55,1.05) circle (2.2pt);
  \fill[removedpt] (1.8,0.95) circle (2.2pt);
  \draw[black!55] (1.55-0.07,1.05-0.07) -- (1.55+0.07,1.05+0.07) (1.55-0.07,1.05+0.07) -- (1.55+0.07,1.05-0.07);
  \draw[black!55] (1.8-0.07,0.95-0.07) -- (1.8+0.07,0.95+0.07) (1.8-0.07,0.95+0.07) -- (1.8+0.07,0.95-0.07);
  \foreach \px/\py in {0.3/2.6, 0.55/1.8, 0.9/1.28, 1.95/0.62, 2.55/0.45, 3.3/0.32} {
    \fill[globalpt] (\px,\py) circle (2pt);
  }
  \filldraw[reportpt] (1.35,0.88) circle (3pt);
  \coordinate (admitted) at (1.35,0.88);
  \draw[discardpt] (2.4,0.9) circle (2.4pt);
  \fill[globalpt] (0.15,-0.6) circle (2pt);   \node[note, anchor=west] at (0.3,-0.6) {global front};
  \filldraw[reportpt] (0.15,-1.0) circle (3pt); \node[note, anchor=west] at (0.3,-1.0) {newly admitted};
  \fill[removedpt] (2.75,-0.6) circle (2.2pt);
  \draw[black!55] (2.75-0.07,-0.6-0.07) -- (2.75+0.07,-0.6+0.07) (2.75-0.07,-0.6+0.07) -- (2.75+0.07,-0.6-0.07);
  \node[note, anchor=west] at (2.9,-0.6) {removed (dominated)};
  \draw[discardpt] (2.75,-1.0) circle (2.4pt); \node[note, anchor=west] at (2.9,-1.0) {newly discarded};
  \node[paneltitle] at (0, 3.3) {Global front $\mathcal{G}$ \textit{(on-chain)}};
\end{scope}
\draw[->] (C.east) -- node[note, below, align=center] {admit or\\discard} (11.3, 0 |- C.east);
\end{tikzpicture}
\caption{Aggregation of the local fronts of multiple solvers into a single global front on-chain. Off-chain, each solver evolves its population and maintains its own local front (filled gray) where the dominated solutions are shown as hollow circles. From the local front, one solution $\mathbf{x}^{\ast}_{t+1}$ (large black) is reported per proof. Each solver submits the objective vector $\mathbf{F}(\mathbf{x}^{\ast}_{t+1})$ together with the proof $\pi_t$ while decision variables always remain private. The smart contract verifies the proof, checks the hash chain and admits the point to the global front $\mathcal{G}$ only if no point of $\mathcal{G}$ dominates it. The points dominated by that newly admitted point are also removed. A submission that is dominated by the global front is discarded. If a front reaches its capacity $K$, it admits the point only if it dominates at least one existing point.}
\label{fig:local-global}
\end{figure*}

In zk-NSGA-II, every solver $i$ maintains its own local front $L_i^t$ off-chain while the contract maintains a single global front $\mathcal{G}$ that aggregates the contributions of all solvers, as shown in Figure \ref{fig:local-global}. In our design, only one solution $\mathbf{x}^{\ast}_{t+1}$ is reported per proof, which leads the global front to be built incrementally. When a verified solution $\mathbf{y} = \mathbf{F}(\mathbf{x}^{\ast}_{t+1})$ is submitted by solver $i$, the contract updates the global front in the following way:
\begin{equation}
\mathcal{G} \leftarrow
\begin{cases}
\mathcal{G}, & \text{if } \exists\, \mathbf{q} \in \mathcal{G} : \mathbf{q} \preceq \mathbf{y} \\
\{\mathbf{q} \in \mathcal{G} : \mathbf{y} \not\preceq \mathbf{q}\} \cup \{\mathbf{y}\}, & \text{otherwise}
\end{cases}
\label{eq:global-update}
\end{equation}
where $\mathbf{q} \preceq \mathbf{y}$ denotes weak dominance. In the first case, the solution is discarded since an existing solution in the global front is at least as good in both objectives. This prevents duplication in the global front as well. In the second case, the solution is admitted and all points it dominates are removed. This update rule preserves that $\mathcal{G}$ is always non-dominated. Since the contract compares each submission against every point in the global front, the gas cost of an update grows linearly with $|\mathcal{G}|$. In this design, we therefore introduce a maximum capacity $K$ for the global front, which bounds this cost by $O(K)$ per submission. When the front is full, a new solution is admitted only if it removes at least one existing point even if that new solution is non-dominated. 

\subsection{Randomness in NSGA-II}
\label{sec:randomness}

NSGA-II is inherently a stochastic algorithm that relies on random numbers for its genetic operators such as the binary tournament selection. However, the guest program of the zkVM has no source of entropy and its execution must be reproducible for verifiability. Therefore, zk-NSGA-II replaces true randomness with a SplitMix64 pseudo-random number generator \cite{steele2014}. The hash chain also provides integrity of the random numbers of every generation and hence, a solver cannot re-run a generation with different random numbers and submit only the most favorable one. Any different random state would generate a different $h_t$ that the contract rejects. However, the seed itself can be chosen freely at the genesis generation just as the initial population. This enables different solvers to have different seeds to improve population diversity. Similarly, the reported solution $\mathbf{x}^{\ast}_{t+1}$ on the local front is drawn from the generator state $s_{t+1}$, which is part of $h_{t+1}$.

\section{Experimental Setup}
\label{sec:setup}

The experiments in this work are specifically designed to evaluate zk-NSGA-II from multiple perspectives. First, zk-NSGA-II is applied to the well-known synthetic benchmark problems with known Pareto fronts to evaluate the correctness of the verifiable execution and the quality of the fronts. Second, it is applied to a real-world portfolio optimization problem to understand the relations of the local fronts with the global front and to measure the cost of verifiability both off-chain and on-chain. Table \ref{tab:setup} provides the key parameters and their values where all experiments are carried out on an Apple MacBook Air M2 with 8 cores and 8 GB RAM. 

\begin{table}[t]
\caption{Experimental setup parameters, grouped by MOO, proof system, blockchain and hardware. These are the default values used in all experiments unless otherwise stated. The gas and ETH prices are used to convert the measured gas consumption into dollars.}
\label{tab:setup}
\small
\begin{tabularx}{\columnwidth}{@{}lX@{}}
\toprule
\textbf{Parameter} & \textbf{Value} \\
\midrule
\multicolumn{2}{@{}l}{\emph{Multi-objective optimization}} \\
Population size $N$ & 50 \\
Number of solvers $S$ & 10 \\
Global front capacity $K$ & 100 \\
Generations $G$ & 250 (ZDT, DTLZ); 1000 (Hang Seng) \\
\midrule
\multicolumn{2}{@{}l}{\emph{Proof system}} \\
zkVM & RISC Zero 3.0.6 \\
Proof type & Groth16 \\
State hash $H(\cdot)$ & SHA-256 \\
\midrule
\multicolumn{2}{@{}l}{\emph{Blockchain}} \\
Smart contracts & Solidity 0.8.26 \\
Proof verifier & RISC Zero Groth16 verifier \\
Gas price & 10 gwei \\
ETH price & 2{,}221 dollars (30 October 2026) \\
Execution environment & Anvil, Foundry 1.8.5 \cite{foundry} \\
\midrule
\multicolumn{2}{@{}l}{\emph{Hardware}} \\
Machine & Apple MacBook Air M2, 8 cores, 8 GB RAM \\
\bottomrule
\end{tabularx}
\end{table}

\subsection{Benchmark Problems}
\label{sec:problems}

We first evaluate the correctness of our zk-NSGA-II implementation on two popular synthetic test suites of ZDT \cite{zitzler2000} and DTLZ \cite{deb2002}, resulting in twelve two-objective problems in total. The problems are scalable in terms of real-valued decision variables with analytically-known Pareto fronts, which allows the quality of the found fronts to be measured exactly against the true front. The problems are chosen to span a broad range of characteristics such as convex, non-convex and disconnected fronts. Complementing these two synthetic benchmarks, we also establish a case study for a real-world portfolio optimization problem in Section \ref{sec:hang-seng}.

\subsubsection{ZDT Suite}

The ZDT problems \cite{zitzler2000} are among the most widely used benchmarks for two-objective optimization. Each problem is constructed from three functions: the first objective $f_1$ depends only on the first decision variable $x_1$; a distance function $g$ controls the proximity to the Pareto front; and the second objective $f_2$ depends on $f_1$ and $g$. The problems in ZDT pose different sorts of difficulty to handle: ZDT1 has a convex Pareto front; ZDT2 has a non-convex front; ZDT3 has a front with five disconnected segments; ZDT4 has the same front as ZDT1 but a multi-modal distance function; and ZDT6 has a non-convex front with a non-uniform density of solutions. ZDT5 is excluded since it is defined over binary decision variables, rather than real-valued variables. The definitions of these five problems are given in Table \ref{tab:zdt} in Appendix \ref{app:problems}.

\subsubsection{DTLZ Suite}

The DTLZ problems \cite{deb2002} are instantiated with two objectives in this work. In each problem, the first decision variable $x_1$ determines the position of a solution along the Pareto front and a distance function $g$ of the remaining variables controls the proximity to the front. As in ZDT, the DTLZ problems pose different sorts of difficulties: DTLZ1 has a linear Pareto front with a multi-modal distance function; DTLZ2 has a non-convex front; DTLZ3 combines the front of DTLZ2 with the multi-modal distance function of DTLZ1; DTLZ4 moves the density of solutions to the extremes of the front; DTLZ6 biases the distance function and DTLZ7 has a disconnected front. DTLZ5 differs from DTLZ2 only in the mapping of the position angles $\theta_2, \dots, \theta_{m-1}$, which do not exist for $m = 2$ objectives. Therefore, DTLZ5 reduces to DTLZ2 in our two-objective setting. We still include DTLZ5 to cover the complete DTLZ1--DTLZ7 suite. The definitions of these seven problems are given in Table \ref{tab:dtlz} in Appendix \ref{app:problems}.

\subsection{Hang Seng Portfolio Problem}
\label{sec:hang-seng}

To evaluate zk-NSGA-II beyond the synthetic benchmarks, we consider the Hang Seng portfolio optimization problem \cite{chang2000} where an investor distributes a total capital into $n$ different assets to minimize risk while maximizing expected return. We believe that this problem is a natural fit for privacy-preserving collaborative optimization since the portfolio weights represent a confidential investment strategy while the risk and return of a portfolio are necessary for comparison. With zk-NSGA-II, the solvers can prove that the reported risk and return belong to a portfolio produced by NSGA-II without disclosing its weights. In the problem, each decision variable $x_j \in [0, 1]$ is mapped to a portfolio weight with normalization:
\begin{equation}
w_j(\mathbf{x}) = \frac{x_j}{\sum_{l=1}^{n} x_l}, \qquad j = 1, \dots, n
\label{eq:portfolio-weights}
\end{equation}
where $w_j \geq 0$ and $\sum_{j} w_j = 1$. Given the vector of mean asset returns $\boldsymbol{\mu}$ and their covariance matrix $\boldsymbol{\Sigma}$, the two objectives are:
\begin{equation}
\begin{aligned}
\min_{\mathbf{x} \in [0,1]^n} \;& f_1(\mathbf{x}) = \mathbf{w}(\mathbf{x})^{\top} \boldsymbol{\Sigma}\, \mathbf{w}(\mathbf{x}) \\
\min_{\mathbf{x} \in [0,1]^n} \;& f_2(\mathbf{x}) = -\boldsymbol{\mu}^{\top} \mathbf{w}(\mathbf{x})
\end{aligned}
\label{eq:portfolio-objectives}
\end{equation}
where $f_1$ is the variance of the portfolio return, which measures its risk and $f_2$ is the negative expected return, so that maximizing the return corresponds to minimizing $f_2$. Both objectives are thus minimized simultaneously.

\subsubsection{Hang Seng Dataset}

The portfolio optimization problem in our work is instantiated with the Hang Seng dataset \cite{chang2000} from the OR-Library \cite{beasley1990}, which is widely used as a benchmark for portfolio optimization. The dataset has 31 assets of the Hang Seng index of Hong Kong, with weekly price data from March 1992 to September 1997 and 291 return observations per asset. For each asset, it provides the mean weekly return $\mu_j$, ranging from $0.000141$ to $0.010865$ and the standard deviation of the return $\sigma_j$, ranging from $0.0358$ to $0.0691$. The dataset includes the efficient frontier of the problem, which serves as the true Pareto front $\mathcal{P}^{\ast}$ for our performance measures.

\subsection{Performance Measures}
\label{sec:measures}

The performance of zk-NSGA-II is measured from three complementary perspectives. For the optimization perspective, four measures are used based on local and global fronts. For the ZKP perspective, four additional measures are considered by relying on only off-chain computation and size. Lastly, for the blockchain perspective, gas consumption of individual contract functions is measured. The performance measures are briefly defined as:

\subsubsection{Number of Non-Dominated Solutions (NS)}

The number of non-dominated solutions counts the solutions in the first non-dominated front \cite{greeff2008}. It is bounded by the population size $N$ for a local front and by the capacity $K$ for the global front. Greater NS refers to a wider set of non-dominated solutions. 

\subsubsection{Schott's Spacing (SS)} Schott's spacing measures how evenly the solutions are distributed on a front \cite{schott1995}. Let $d_i$ be the Manhattan distance from the $i$th solution of a front $\mathcal{P}$ to its nearest neighbor in objective space. Then, the spacing is:
\begin{equation}
\mathrm{SS}(\mathcal{P}) = \sqrt{\frac{1}{|\mathcal{P}| - 1} \sum_{i=1}^{|\mathcal{P}|} \left(\bar{d} - d_i\right)^2 }
\label{eq:ss}
\end{equation}
where $\bar{d}$ is the mean distance. Lower SS values mean better solution distribution on a front. 

\subsubsection{Inverted Generational Distance (IGD)} The inverted generational distance takes the average distance from each point of a true front $\mathcal{P}^{\ast}$ to the nearest solution of a found front $\mathcal{P}$ \cite{coello2004}:
\begin{equation}
\mathrm{IGD}(\mathcal{P}, \mathcal{P}^{\ast}) = \frac{1}{|\mathcal{P}^{\ast}|} \sum_{\mathbf{z} \in \mathcal{P}^{\ast}} \min_{\mathbf{p} \in \mathcal{P}} \lVert \mathbf{z} - \mathbf{p} \rVert_2
\label{eq:igd}
\end{equation}
where $\mathbf{p}$ refers to an individual solution on $\mathcal{P}$. This measure requires the true front to be known in advance. In our work, both synthetic problems and the portfolio problem satisfy this requirement. 

\subsubsection{Hypervolume Ratio (HVR)} The hypervolume ratio (HVR) normalizes the hypervolume of a found front $\mathcal{P}$ with the hypervolume of the true front $\mathcal{P}^{\ast}$ \cite{zitzler1999}:
\begin{equation}
\begin{gathered}
\mathrm{HVR}(\mathcal{P}) = \frac{\mathrm{HV}(\mathcal{P}, \mathbf{r})}{\mathrm{HV}(\mathcal{P}^{\ast}, \mathbf{r})} \\
\mathrm{HV}(A, \mathbf{r}) = \lambda\left( \bigcup_{\mathbf{p} \in A} \{ \mathbf{y} \in \mathbb{R}^2 : \mathbf{p} \preceq \mathbf{y} \preceq \mathbf{r} \} \right)
\end{gathered}
\label{eq:hv}
\end{equation}
where $\mathrm{HV}(A, \mathbf{r})$ is the area of the objective space between the set $A$ and the reference point $\mathbf{r}$ while $\lambda$ denotes the Lebesgue measure. A higher HVR is better where a value of one means that the found front covers the same area as the true front. In our implementation, the reference point is placed 10\% beyond the nadir point, which is the point of the worst objective values on the true front.

\subsubsection{zkVM Cycles} The computational cost of a proof is measured with the number of cycles that zkVM needs to execute. For this metric, cycles are first measured for a single generation proof and then reported after the decomposition into individual evolutionary phases such as non-dominated sorting.

\subsubsection{Proving Time}

The proving time is the wall-clock time to generate a Groth16 proof for a single generation, including the internal STARK proving of the execution and its compression into Groth16. Although the RISC Zero zkVM provides three alternative proof types as composite, succinct and Groth16 \cite{bruestle2023}, this work considers only Groth16 since it is the only proof type that can be verified on-chain at a constant cost. 

\subsubsection{Verification Time}

The verification time is the wall-clock time to verify a proof off-chain. It is reported as the average of multiple consecutive verifications. Note that on-chain verification is evaluated differently through blockchain gas consumption.

\subsubsection{Proof Size} The proof size is the size of the Groth16 proof in bytes, which is submitted to the blockchain together with the public outputs of the off-chain computation. Since a Groth16 proof consists of a fixed number of elliptic curve points, its size is constant regardless of the population size, the number of decision variables and the length of the execution. 

\subsubsection{Blockchain Gas Consumption} Gas consumption measures the on-chain gas costs of smart contract functions. It is measured from the receipts of transactions executed on a local Ethereum blockchain. Since gas prices and exchange rates fluctuate constantly, gas consumption is converted into ETH and US dollars with 10 gwei per unit of gas and an ETH price of 2{,}221 US dollars as of 30 October 2026. The cost of a function is therefore its gas consumption multiplied by the gas price and the ETH price.

\section{Experimental Results}
\label{sec:results}

This section reports the results of the experiments. Section \ref{sec:res-quality} first validates that zk-NSGA-II produces correct results on the ZDT and DTLZ problems with known fronts. Section \ref{sec:res-local-global} examines the aggregation of the local fronts of the solvers into the global front on-chain. Sections \ref{sec:res-proving} and \ref{sec:res-onchain} quantify the cost of verifiability both off-chain and on-chain. Finally, Section \ref{sec:res-portfolio} presents the results of zk-NSGA-II on the portfolio optimization case study. 

\subsection{Performance on ZDT and DTLZ Problems}
\label{sec:res-quality}

This section evaluates zk-NSGA-II on the ZDT and DTLZ problems with known Pareto fronts. For this experiment, we execute NSGA-II as the guest program natively.

\begin{figure*}[t]
\centering
\includegraphics[width=0.75\textwidth]{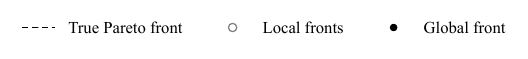}\\[2pt]
\begin{subfigure}[t]{0.245\textwidth}\centering\includegraphics[width=\linewidth]{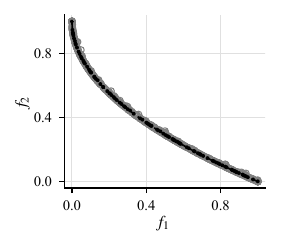}\caption{ZDT1}\label{fig:fronts-snapshots-zdt1}\end{subfigure}
\hfill
\begin{subfigure}[t]{0.245\textwidth}\centering\includegraphics[width=\linewidth]{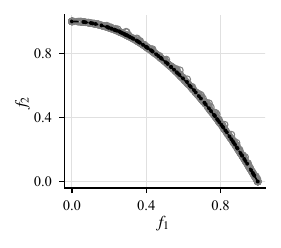}\caption{ZDT2}\label{fig:fronts-snapshots-zdt2}\end{subfigure}
\hfill
\begin{subfigure}[t]{0.245\textwidth}\centering\includegraphics[width=\linewidth]{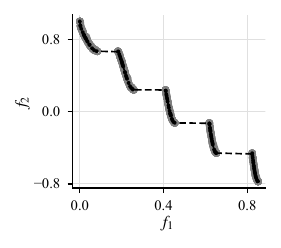}\caption{ZDT3}\label{fig:fronts-snapshots-zdt3}\end{subfigure}
\hfill
\begin{subfigure}[t]{0.245\textwidth}\centering\includegraphics[width=\linewidth]{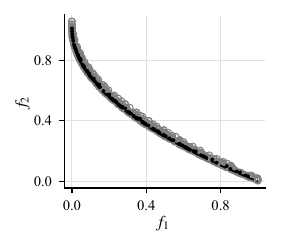}\caption{ZDT4}\label{fig:fronts-snapshots-zdt4}\end{subfigure}
\\[2pt]
\begin{subfigure}[t]{0.245\textwidth}\centering\includegraphics[width=\linewidth]{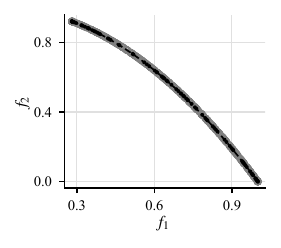}\caption{ZDT6}\label{fig:fronts-snapshots-zdt6}\end{subfigure}
\hfill
\begin{subfigure}[t]{0.245\textwidth}\centering\includegraphics[width=\linewidth]{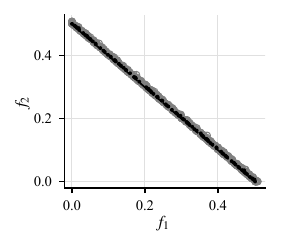}\caption{DTLZ1}\label{fig:fronts-snapshots-dtlz1}\end{subfigure}
\hfill
\begin{subfigure}[t]{0.245\textwidth}\centering\includegraphics[width=\linewidth]{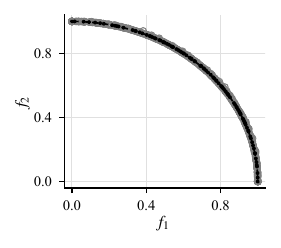}\caption{DTLZ2}\label{fig:fronts-snapshots-dtlz2}\end{subfigure}
\hfill
\begin{subfigure}[t]{0.245\textwidth}\centering\includegraphics[width=\linewidth]{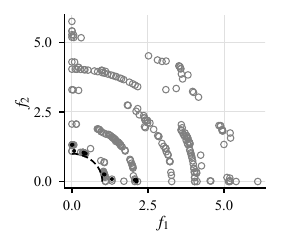}\caption{DTLZ3}\label{fig:fronts-snapshots-dtlz3}\end{subfigure}
\\[2pt]
\begin{subfigure}[t]{0.245\textwidth}\centering\includegraphics[width=\linewidth]{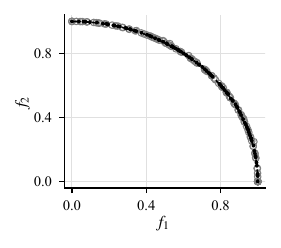}\caption{DTLZ4}\label{fig:fronts-snapshots-dtlz4}\end{subfigure}
\hfill
\begin{subfigure}[t]{0.245\textwidth}\centering\includegraphics[width=\linewidth]{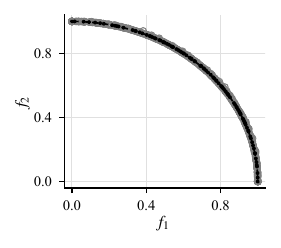}\caption{DTLZ5}\label{fig:fronts-snapshots-dtlz5}\end{subfigure}
\hfill
\begin{subfigure}[t]{0.245\textwidth}\centering\includegraphics[width=\linewidth]{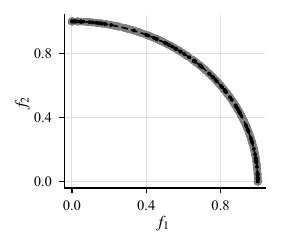}\caption{DTLZ6}\label{fig:fronts-snapshots-dtlz6}\end{subfigure}
\hfill
\begin{subfigure}[t]{0.245\textwidth}\centering\includegraphics[width=\linewidth]{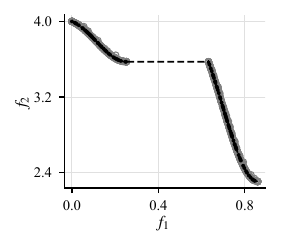}\caption{DTLZ7}\label{fig:fronts-snapshots-dtlz7}\end{subfigure}
\\[2pt]
\caption{Final fronts of zk-NSGA-II at the last generation $G = 250$ for the twelve ZDT and DTLZ problems, each with $S = 10$ solvers of population size $N = 50$ and a global front capacity of $K = 100$. Each panel plots the two objectives $f_1$ and $f_2$ and overlays the known true Pareto front, the local fronts of the solvers kept off-chain (hollow gray circles) and the global front on-chain from their submissions (filled black circles).}
\label{fig:res-fronts}
\end{figure*}

\textbf{The local fronts converge to the true Pareto fronts on all problems except DTLZ3 and DTLZ4.} As shown in Table \ref{tab:res-quality}, the local fronts reach mean HVR values between $0.9511$ and $0.9938$ on the remaining ten problems. On eight of them, the mean IGD stays between $0.0066$ and $0.0125$ with the HVR standard deviation of at most $0.0082$. The two exceptions are ZDT4 and DTLZ7 with the high standard deviations of IGD ($0.0287 \pm 0.0193$ and $0.0544 \pm 0.1372$) and HVR ($0.9511 \pm 0.0255$ and $0.9636 \pm 0.0851$). The number of non-dominated solutions reaches $50$ on all ten problems except ZDT4 ($47.5 \pm 7.9$). Schott's spacing is between $0.0093$ and $0.0196$, which indicates that the solutions are evenly spread along the fronts, as shown in Figure \ref{fig:res-fronts}. Note that DTLZ5 yields the same values as DTLZ2 since the two problems have the same two objectives. This experiment validates the correctness of zk-NSGA-II on the zkVM.

\textbf{DTLZ3 and DTLZ4 remain hard to solve due to the known difficulties of these problems.} In DTLZ3, the multi-modal function traps the populations in local fronts with the mean IGD of $2.4067 \pm 1.4631$ and the HVR of $0.0422 \pm 0.0966$. In DTLZ4, the density bias maps most values of $x_1$ to the extremes of the front with the mean IGD of $0.5989 \pm 0.3097 $ and the HVR of $0.4023 \pm 0.3015 $. These high standard deviations imply that the results of individual problem solvers depend on the initial population and the random seed since they all execute the same zk-NSGA-II program. These problems also motivate further collaboration among solvers such as privacy-preserving solution sharing. 

\begin{figure*}[t]
\centering
\includegraphics[width=0.75\textwidth]{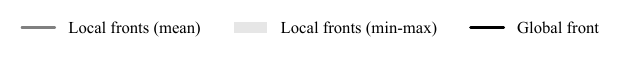}\\[2pt]
\begin{subfigure}[t]{0.245\textwidth}\centering\includegraphics[width=\linewidth]{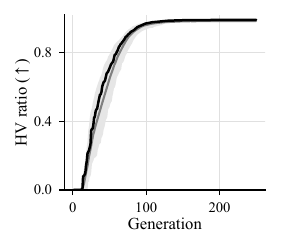}\caption{ZDT1}\label{fig:res-conv-hv-zdt1}\end{subfigure}
\hfill
\begin{subfigure}[t]{0.245\textwidth}\centering\includegraphics[width=\linewidth]{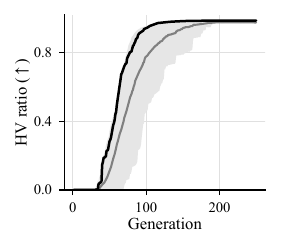}\caption{ZDT2}\label{fig:res-conv-hv-zdt2}\end{subfigure}
\hfill
\begin{subfigure}[t]{0.245\textwidth}\centering\includegraphics[width=\linewidth]{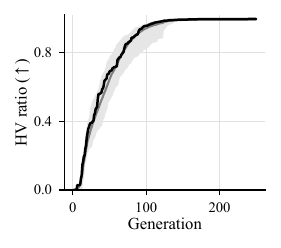}\caption{ZDT3}\label{fig:res-conv-hv-zdt3}\end{subfigure}
\hfill
\begin{subfigure}[t]{0.245\textwidth}\centering\includegraphics[width=\linewidth]{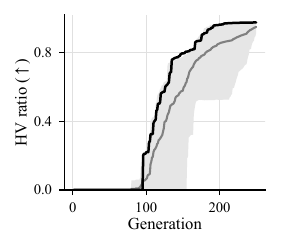}\caption{ZDT4}\label{fig:res-conv-hv-zdt4}\end{subfigure}
\\[2pt]
\begin{subfigure}[t]{0.245\textwidth}\centering\includegraphics[width=\linewidth]{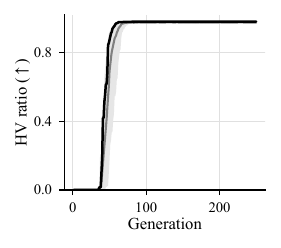}\caption{ZDT6}\label{fig:res-conv-hv-zdt6}\end{subfigure}
\hfill
\begin{subfigure}[t]{0.245\textwidth}\centering\includegraphics[width=\linewidth]{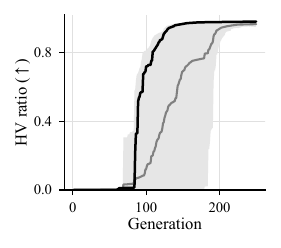}\caption{DTLZ1}\label{fig:res-conv-hv-dtlz1}\end{subfigure}
\hfill
\begin{subfigure}[t]{0.245\textwidth}\centering\includegraphics[width=\linewidth]{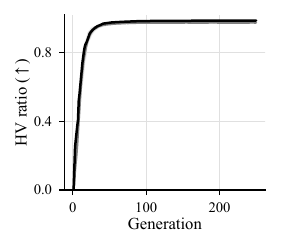}\caption{DTLZ2}\label{fig:res-conv-hv-dtlz2}\end{subfigure}
\hfill
\begin{subfigure}[t]{0.245\textwidth}\centering\includegraphics[width=\linewidth]{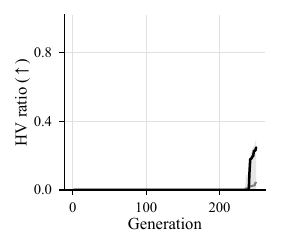}\caption{DTLZ3}\label{fig:res-conv-hv-dtlz3}\end{subfigure}
\\[2pt]
\begin{subfigure}[t]{0.245\textwidth}\centering\includegraphics[width=\linewidth]{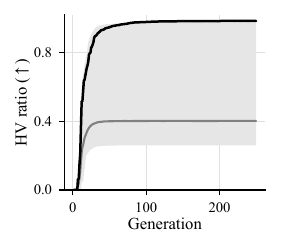}\caption{DTLZ4}\label{fig:res-conv-hv-dtlz4}\end{subfigure}
\hfill
\begin{subfigure}[t]{0.245\textwidth}\centering\includegraphics[width=\linewidth]{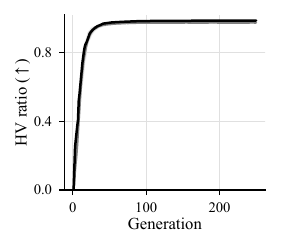}\caption{DTLZ5}\label{fig:res-conv-hv-dtlz5}\end{subfigure}
\hfill
\begin{subfigure}[t]{0.245\textwidth}\centering\includegraphics[width=\linewidth]{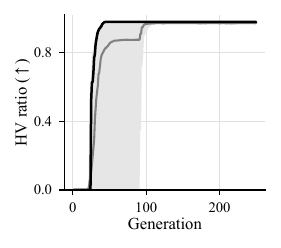}\caption{DTLZ6}\label{fig:res-conv-hv-dtlz6}\end{subfigure}
\hfill
\begin{subfigure}[t]{0.245\textwidth}\centering\includegraphics[width=\linewidth]{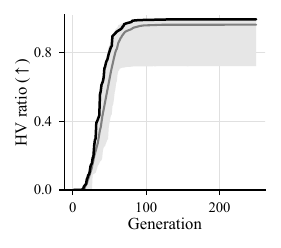}\caption{DTLZ7}\label{fig:res-conv-hv-dtlz7}\end{subfigure}
\\[2pt]
\caption{Convergence of the HVR over the $G = 250$ generations for the twelve ZDT and DTLZ problems with $S = 10$ solvers. The black line is the HVR of the on-chain global front, the gray line is the mean HVR of the 10 off-chain local fronts and the shaded area spans the minimum and maximum HVR among the solvers at each generation.}
\label{fig:res-convergence}
\end{figure*}

\begin{figure*}[t]
\centering
\includegraphics[width=0.75\textwidth]{figures/results/legend_convergence}\\[2pt]
\begin{subfigure}[t]{0.245\textwidth}\centering\includegraphics[width=\linewidth]{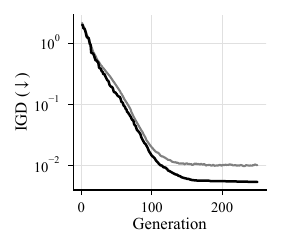}\caption{ZDT1}\label{fig:res-conv-igd-zdt1}\end{subfigure}
\hfill
\begin{subfigure}[t]{0.245\textwidth}\centering\includegraphics[width=\linewidth]{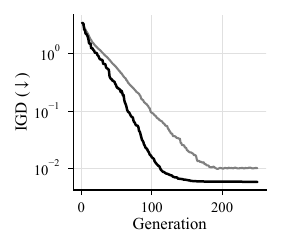}\caption{ZDT2}\label{fig:res-conv-igd-zdt2}\end{subfigure}
\hfill
\begin{subfigure}[t]{0.245\textwidth}\centering\includegraphics[width=\linewidth]{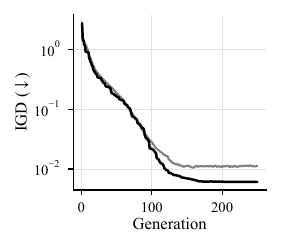}\caption{ZDT3}\label{fig:res-conv-igd-zdt3}\end{subfigure}
\hfill
\begin{subfigure}[t]{0.245\textwidth}\centering\includegraphics[width=\linewidth]{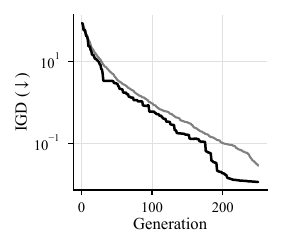}\caption{ZDT4}\label{fig:res-conv-igd-zdt4}\end{subfigure}
\\[2pt]
\begin{subfigure}[t]{0.245\textwidth}\centering\includegraphics[width=\linewidth]{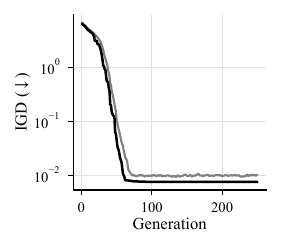}\caption{ZDT6}\label{fig:res-conv-igd-zdt6}\end{subfigure}
\hfill
\begin{subfigure}[t]{0.245\textwidth}\centering\includegraphics[width=\linewidth]{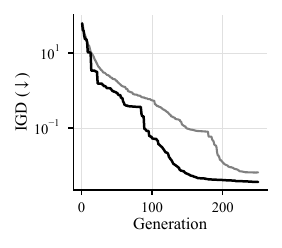}\caption{DTLZ1}\label{fig:res-conv-igd-dtlz1}\end{subfigure}
\hfill
\begin{subfigure}[t]{0.245\textwidth}\centering\includegraphics[width=\linewidth]{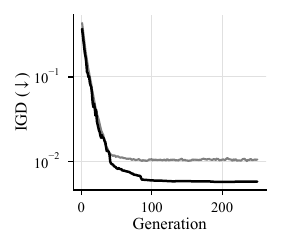}\caption{DTLZ2}\label{fig:res-conv-igd-dtlz2}\end{subfigure}
\hfill
\begin{subfigure}[t]{0.245\textwidth}\centering\includegraphics[width=\linewidth]{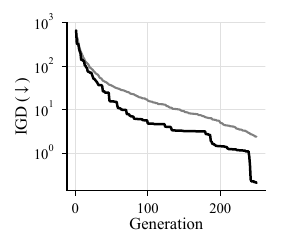}\caption{DTLZ3}\label{fig:res-conv-igd-dtlz3}\end{subfigure}
\\[2pt]
\begin{subfigure}[t]{0.245\textwidth}\centering\includegraphics[width=\linewidth]{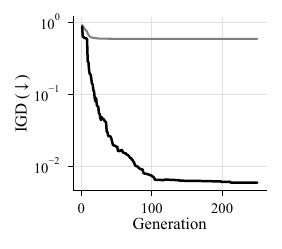}\caption{DTLZ4}\label{fig:res-conv-igd-dtlz4}\end{subfigure}
\hfill
\begin{subfigure}[t]{0.245\textwidth}\centering\includegraphics[width=\linewidth]{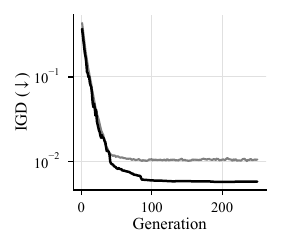}\caption{DTLZ5}\label{fig:res-conv-igd-dtlz5}\end{subfigure}
\hfill
\begin{subfigure}[t]{0.245\textwidth}\centering\includegraphics[width=\linewidth]{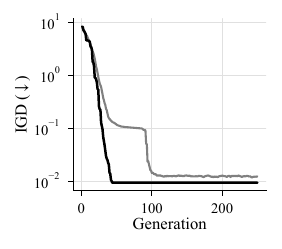}\caption{DTLZ6}\label{fig:res-conv-igd-dtlz6}\end{subfigure}
\hfill
\begin{subfigure}[t]{0.245\textwidth}\centering\includegraphics[width=\linewidth]{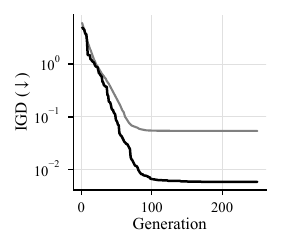}\caption{DTLZ7}\label{fig:res-conv-igd-dtlz7}\end{subfigure}
\\[2pt]
\caption{Convergence of the IGD over the $G = 250$ generations for the twelve ZDT and DTLZ problems with $S = 10$ solvers. The black line is the IGD of the on-chain global front and the gray line is the mean IGD of the 10 off-chain local fronts.}
\label{fig:res-convergence-igd}
\end{figure*}

\textbf{The global front converges faster than the average solver.} Figures \ref{fig:res-convergence} and \ref{fig:res-convergence-igd} show how the global front and the local fronts evolve over the generations for the twelve synthetic problems. Figure \ref{fig:res-convergence} plots the HVR of the global front together with the mean HVR of the local fronts while the shaded area indicates the minimum and the maximum HVR bounds among all the solvers. Figure \ref{fig:res-convergence-igd} plots the IGD of the global front and the mean IGD of the local fronts. On every problem, it is observed that the global front is at least as good as the mean local front in both HVR and IGD in most of the generations. For instance, the global front reaches an HVR of $0.9$ on ZDT2 at generation $89$ while the mean local front needs $130$ generations. Similarly, it needs $124$ versus $193$ generations on DTLZ1, $176$ versus $230$ on ZDT4, and $33$ versus $92$ on DTLZ6. This favors the problem owner who gets a front close to the best solver at every generation without knowing who will perform best.

\begin{table*}[t]
\caption{Quality of the local fronts and the global front at generation $G = 250$ for the ZDT and DTLZ problems. Each problem is solved by $S = 10$ solvers with population size $N = 50$ and a global front capacity of $K = 100$. Arrows indicate whether higher ($\uparrow$) or lower ($\downarrow$) values are better. The acceptance rate is the share of submissions admitted to the global front.}
\label{tab:res-quality}
\footnotesize
\setlength{\tabcolsep}{2.5pt}
\begin{tabular*}{\textwidth}{@{\extracolsep{\fill}}lccccccccccc@{}}
\toprule
& \multicolumn{5}{c}{\textbf{Local fronts}} & \multicolumn{3}{c}{\textbf{Global front}} & \multicolumn{3}{c}{\textbf{Acceptance rate (\%)}} \\
\cmidrule(lr){2-6} \cmidrule(lr){7-9} \cmidrule(l){10-12}
\textbf{Problem} & \textbf{NS} $\uparrow$ & \textbf{SS} $\downarrow$ & \textbf{IGD} $\downarrow$ & \textbf{HVR} $\uparrow$ & \textbf{Max HVR} $\uparrow$ & \textbf{NS} & \textbf{IGD} $\downarrow$ & \textbf{HVR} $\uparrow$ & \textbf{All} & \textbf{First 25} & \textbf{Last 25} \\
\midrule
ZDT1 & $50.0 \pm 0.0$ & $0.0163 \pm 0.0021$ & $0.0100 \pm 0.0006$ & $0.9853 \pm 0.0009$ & 0.9869 & 100 & 0.0053 & 0.9922 & 30.4 & 50.0 & 5.6 \\
ZDT2 & $50.0 \pm 0.0$ & $0.0149 \pm 0.0021$ & $0.0102 \pm 0.0007$ & $0.9763 \pm 0.0020$ & 0.9790 & 100 & 0.0059 & 0.9879 & 22.7 & 34.8 & 4.8 \\
ZDT3 & $50.0 \pm 0.0$ & $0.0196 \pm 0.0018$ & $0.0114 \pm 0.0004$ & $0.9938 \pm 0.0004$ & 0.9947 & 100 & 0.0061 & 0.9968 & 37.2 & 53.2 & 3.6 \\
ZDT4 & $47.5 \pm 7.9$ & $0.0187 \pm 0.0088$ & $0.0287 \pm 0.0193$ & $0.9511 \pm 0.0255$ & 0.9769 & 100 & 0.0115 & 0.9766 & 22.5 & 21.6 & 12.4 \\
ZDT6 & $50.0 \pm 0.0$ & $0.0171 \pm 0.0008$ & $0.0103 \pm 0.0007$ & $0.9750 \pm 0.0015$ & 0.9778 & 100 & 0.0076 & 0.9818 & 12.2 & 45.6 & 0.0 \\
DTLZ1 & $50.0 \pm 0.0$ & $0.0093 \pm 0.0010$ & $0.0066 \pm 0.0014$ & $0.9667 \pm 0.0082$ & 0.9746 & 100 & 0.0037 & 0.9816 & 19.8 & 34.4 & 6.4 \\
DTLZ2 & $50.0 \pm 0.0$ & $0.0169 \pm 0.0026$ & $0.0106 \pm 0.0005$ & $0.9752 \pm 0.0016$ & 0.9771 & 100 & 0.0058 & 0.9878 & 21.5 & 62.4 & 5.6 \\
DTLZ3 & $26.0 \pm 15.1$ & $0.2966 \pm 0.2958$ & $2.4067 \pm 1.4631$ & $0.0422 \pm 0.0966$ & 0.2910 & 13 & 0.2071 & 0.2475 & 13.5 & 33.6 & 19.2 \\
DTLZ4 & $11.2 \pm 20.5$ & $0.0056 \pm 0.0087$ & $0.5989 \pm 0.3097$ & $0.4023 \pm 0.3015$ & 0.9755 & 100 & 0.0060 & 0.9854 & 13.6 & 24.0 & 2.8 \\
DTLZ5 & $50.0 \pm 0.0$ & $0.0169 \pm 0.0026$ & $0.0106 \pm 0.0005$ & $0.9752 \pm 0.0016$ & 0.9771 & 100 & 0.0058 & 0.9878 & 21.5 & 62.4 & 5.6 \\
DTLZ6 & $50.0 \pm 0.0$ & $0.0194 \pm 0.0023$ & $0.0125 \pm 0.0005$ & $0.9721 \pm 0.0013$ & 0.9750 & 100 & 0.0095 & 0.9793 & 9.0 & 49.6 & 0.0 \\
DTLZ7 & $50.0 \pm 0.0$ & $0.0159 \pm 0.0038$ & $0.0544 \pm 0.1372$ & $0.9636 \pm 0.0851$ & 0.9913 & 100 & 0.0059 & 0.9950 & 24.8 & 46.0 & 4.4 \\
\bottomrule
\end{tabular*}
\end{table*}

\subsection{Aggregation of Local Fronts to Global Front}
\label{sec:res-local-global}

This section discusses the empirical findings of aggregating multiple local fronts off-chain into a single global front on-chain. 

\textbf{The aggregation compensates for the solvers that fail.} The global front outperforms the best local front on ten of the twelve problems. The global front reaches an HVR of $0.9922$ on ZDT1 and $0.9950$ on DTLZ7 while the best local fronts reach $0.9869$ and $0.9913$ in Table \ref{tab:res-quality}. The first of the exceptions is ZDT4 where the global front ($0.9766$) and the best local front ($0.9769$) are nearly equal. The second one is DTLZ3 where the global front ($0.2475$) exceeds the mean local front ($0.0422$) but falls short of the best local front ($0.2910$). The solvers start from populations at different regions of the available search space. Not all the solvers are lucky enough to start from or reach promising regions. Hence, the front aggregation compensates for these solvers by using the well-performing ones. Currently, nothing prevents independent solvers from exploring overlapping regions, which might partially waste their effort. Such coordination among the solvers is left for future work.

\textbf{The acceptance rate decreases over generations.} The acceptance rate of submissions to the global front is between $21.6\%$ and $62.4\%$ in the first $25$ generations and between $0.0\%$ and $19.2\%$ in the last $25$ generations in Table \ref{tab:res-quality}. ZDT6 and DTLZ6 admit no submission at all in the last $25$ generations. At the early generations, the global front is sparse where most submissions are likely to be non-dominated. However, at the later generations, the global front is full and close to the true front where solutions are admitted only if they dominate the existing ones. Therefore, especially the problems that are still converging keep the highest late acceptance such as DTLZ3 with $19.2\%$ and ZDT4 with $12.4\%$. This builds an interesting connection between the acceptance rate and the convergence of the global front. The acceptance rate and the global front size over generations are given in Figures \ref{fig:res-acceptance} and \ref{fig:res-size} in Appendix \ref{app:acceptance}. In addition, this might motivate early solution submissions and discourage solvers to join late. 

\subsection{Off-Chain Proof Generation Cost}
\label{sec:res-proving}

This section presents the empirical findings of generating a single zk-NSGA-II proof off-chain. Note that the zkVM cycles in this experiment count the RISC-V instructions that the guest program $\Gamma$ executes inside the zkVM (i.e., cost of proving) while the native CPU time measures the same generation directly on the host CPU without proving (i.e., cost of computation itself).

\begin{figure}[t]
\centering
\includegraphics[width=\linewidth]{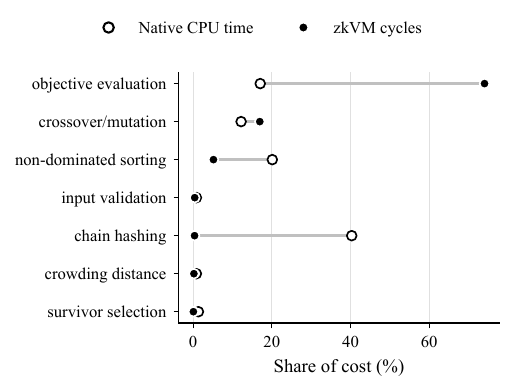}
\caption{Share of the zkVM cycles and the native CPU time of each algorithm component in a single proven generation of the Hang Seng portfolio problem. Filled circles give the share of zkVM cycles (i.e., the cost of proving) and hollow circles give the share of native CPU time on the host without proving (i.e., the cost of the computation itself).}
\label{fig:res-profile}
\end{figure}

\textbf{Objective evaluation dominates the proof generation cost in the zkVM while NSGA-II-native genetic operators remain relatively cheap.} For the Hang Seng portfolio problem shown in Figure \ref{fig:res-profile}, objective evaluation accounts for $74.07\%$ of the zkVM cycles but only $17.05\%$ of the native CPU time. On the other hand, non-dominated sorting decreases from $20.12\%$ of the native time to $5.16\%$ of the cycles and chain hashing from $40.31\%$ to $0.36\%$. While crossover/mutation takes $16.94\%$ of the cycles, crowding distance and survivor selection together stay under $1\%$. The issue here is that the zkVM has no support for floating-point unit, which therefore must be emulated over many cycles. In this regard, the quadratic form of the portfolio variance with multiple multiplications and additions results in most of the cycles. Non-dominated sorting performs only comparisons, which is cheaper. Overall, the objective functions of the problem being solved are of critical importance in terms of zkVM cycles.  

\textbf{Generating a single proof takes about half an hour, whereas verifying it takes only milliseconds regardless of computation.} As Table \ref{tab:res-proving} shows, a ZDT1 generation executes in $8.43 \times 10^{6}$ cycles with $9.44 \times 10^{6}$ proving cycles. Generating a single Groth16 proof takes $1{,}947.8$\,s in total while verifying the STARK and Groth16 proofs off-chain takes only $11.61$\,ms and $2.84$\,ms, respectively. Note that the Groth16 proof here makes the on-chain verification very affordable since the STARK proof is around $222{,}668$\,B while the Groth16 proof is just $260$\,B. In addition, hash-heavy verification of the STARK proof is too expensive for the EVM. Compression to Groth16 therefore enables the verifier contract to accept and verify this succinct proof on-chain. Overall, despite the expensive computation needed to be executed by the prover, the verifier is responsible for checking only a constant-size proof.

\begin{table}[t]
\caption{Proof generation and off-chain verification of a single generation of ZDT1. Proving is split into the STARK proving of the execution and compression into a Groth16 proof.}
\label{tab:res-proving}
\small
\begin{tabularx}{\columnwidth}{@{}Xr@{}}
\toprule
\textbf{Measure} & \textbf{Value} \\
\midrule
\multicolumn{2}{@{}l}{\emph{Execution}} \\
zkVM cycles & $8.43 \times 10^{6}$ \\
Proving cycles & $9.44 \times 10^{6}$ \\
\midrule
\multicolumn{2}{@{}l}{\emph{Proof generation}} \\
STARK proving to succinct receipt & 1{,}763.5\,s \\
Compression to Groth16 & 184.3\,s \\
Total proving time & 1{,}947.8\,s \\
\midrule
\multicolumn{2}{@{}l}{\emph{Proof size}} \\
Succinct STARK proof & 222{,}668\,B \\
Groth16 proof & 260\,B \\
\midrule
\multicolumn{2}{@{}l}{\emph{Off-chain verification}} \\
Succinct STARK proof & $11.61$\,ms \\
Groth16 proof & $2.84$\,ms \\
\bottomrule
\end{tabularx}
\end{table}

\subsection{On-Chain Proof Verification Cost}
\label{sec:res-onchain}

This section discusses the results of verifying a zk-NSGA-II proof on-chain. The gas consumption is measured from transaction receipts on a local Ethereum node. The conversion rates from gas units to ETH and USD are set to 10 gwei and 2{,}221 dollars, respectively. 

\textbf{Proof submission takes a fixed gas consumption regardless of the NSGA-II computation behind the proof.} As shown in Table \ref{tab:res-gas}, a \texttt{submit} transaction for ZDT1 consumes $348{,}961$ gas (around $0.0035$ ETH and $7.75$ dollars). The Groth16 verification alone consumes $250{,}960$ gas, accounting for $71.9\%$ of the submission. The remaining $98{,}001$ gas consumption includes additional checks such as for hash chain or admission to the global front. For the sake of simplicity and reproducibility, we intentionally measure consumption on an empty global front to determine the lower bound. On the global front with multiple non-dominated solutions, the consumption might increase due to the removal of dominated solutions from the front. Overall, the proof system rather than the optimization problem is the critical factor for the on-chain cost of verifiability. This implies that the harder problems might be more expensive to prove off-chain but not to verify on-chain.

\textbf{Deployment is a one-time cost and reading the global front is free.} Deploying the Groth16 Verifier and the Pareto Front contract consumes $1{,}179{,}986$ and $720{,}797$ gas, respectively. The problem owner needs to pay around $42.22$ dollars in total per problem. Reading the global front through \texttt{getFront} is an off-chain call that consumes no gas at all. However, the $28{,}322$ gas is necessary when another contract reads it within another transaction. From a system perspective, suppose a scenario with $S = 10$ solvers for $G = 250$ generations. A ZDT1 run results in $2{,}500$ submissions with around $1{,}940$ dollars per solver, which is quite expensive for most practical applications. Here, the number of generations a solver participates in determines the main cost of on-chain verifiability.

\begin{table*}[t]
\caption{Gas consumption of the contract functions for ZDT1, converted with a gas price of 10 gwei and an ETH price of 2{,}221 dollars (as of 30 October 2026).}
\label{tab:res-gas}
\small
\setlength{\tabcolsep}{4pt}
\begin{tabularx}{\textwidth}{@{}Xrrr@{}}
\toprule
\textbf{Function} & \textbf{Gas} & \textbf{ETH} & \textbf{USD} \\
\midrule
\multicolumn{4}{@{}l}{\emph{Deployment by the problem owner (once per problem)}} \\
Groth16 verifier contract & 1{,}179{,}986 & 0.0118 & 26.21 \\
Pareto Front contract & 720{,}797 & 0.0072 & 16.01 \\
\midrule
\multicolumn{4}{@{}l}{\emph{Submission by a solver (once per generation)}} \\
\texttt{submit}: proof verification, solver and hash-chain checks and global front update & 348{,}961 & 0.0035 & 7.75 \\
\texttt{verify}: Groth16 proof verification only & 250{,}960 & 0.0025 & 5.57 \\
\midrule
\multicolumn{4}{@{}l}{\emph{Reading the global front (free as an off-chain call)}} \\
\texttt{getFront}: all solutions of $\mathcal{G}$ & 28{,}322 & 0.0003 & 0.63 \\
\texttt{size}: number of solutions in $\mathcal{G}$ & 23{,}387 & 0.0002 & 0.52 \\
\bottomrule
\end{tabularx}
\end{table*}

\subsection{Portfolio Case Study}
\label{sec:res-portfolio}

In this experiment, we apply zk-NSGA-II to the real-world problem of the Hang Seng portfolio optimization with $S = 10$ solvers, $N = 50$ portfolios per solver and $G = 1000$ generations. 

\begin{figure*}[t]
\centering
\includegraphics[width=0.82\textwidth]{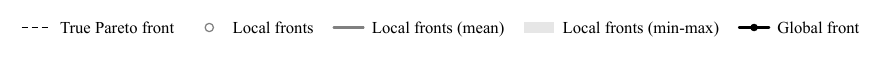}\\[2pt]
\begin{subfigure}[t]{0.32\textwidth}\centering\includegraphics[width=0.92\linewidth]{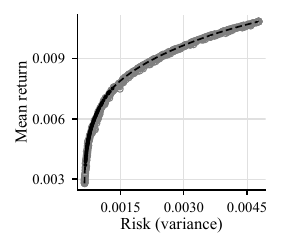}\caption{Fronts at generation 1000}\label{fig:res-portfolio-fronts}\end{subfigure}\hfill
\begin{subfigure}[t]{0.32\textwidth}\centering\includegraphics[width=0.92\linewidth]{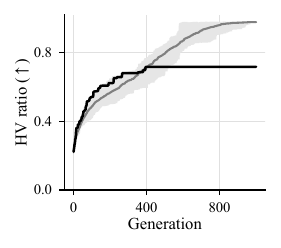}\caption{HVR}\label{fig:res-portfolio-hv}\end{subfigure}\hfill
\begin{subfigure}[t]{0.32\textwidth}\centering\includegraphics[width=0.92\linewidth]{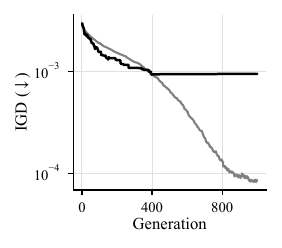}\caption{IGD}\label{fig:res-portfolio-igd}\end{subfigure}
\caption{Results of zk-NSGA-II on the Hang Seng portfolio problem with $n = 31$ assets, $S = 10$ solvers, $N = 50$ portfolios per solver, $G = 1000$ generations and a global front capacity of $K = 100$: (a) the minimization of risk versus the maximization of mean return, (b) HVR and (c) IGD over generations for the global front.}
\label{fig:res-portfolio}
\end{figure*}

\textbf{The global front outperforms the local fronts at the early generations but stalls once it reaches its capacity.} As shown in Figure \ref{fig:res-portfolio}, the global front reaches an HVR of $0.542$ at generation $100$ and $0.658$ at generation $250$ while the mean local front reaches $0.486$ and $0.595$, respectively. However, the global front fills its capacity of $K = 100$ solutions at generation $198$ and its HVR stalls at $0.718$ after generation $500$. The local fronts keep converging to the HVR of $0.978$ at generation $1000$. With a similar trend, the IGD of the global front ends at $9.54 \times 10^{-4}$ while the mean IGD of the local fronts falls to $8.5 \times 10^{-5}$. Recall that once the global front is full, the contract admits a new solution only if it removes at least one dominated solution in our design. Portfolios that extend the front to the higher risk and the higher return are essentially non-dominated but dominate no solutions at the global front, resulting in being discarded without admission. Of the $8{,}030$ submissions from generation $198$ on, $4{,}147$ (i.e., $51.6\%$) are discarded because the global front is full and only $276$ are admitted. This has two implications. First, the capacity favors the first-come admission. Second, the capacity introduces a trade-off between efficiency in gas consumption (since on-chain solution comparison would increase otherwise) and global front quality.

\textbf{The problem owner receives a verified but incomplete frontier.} Each problem solver evolves its own population, converging to the true efficient frontier as in Figure \ref{fig:res-portfolio-fronts} while it is not true for the global front. Over all $10{,}000$ submissions, only $849$ ($8.5\%$) are admitted and all admissions are verified for an honest NSGA-II execution against solution fabrication. This differs from the ZDT and DTLZ problems (Section \ref{sec:res-local-global}) since the global front fills its capacity in $11$ of the $12$ problems and their true fronts are covered well by $K = 100$ solutions with $G = 250$ generations. Hence, careful selection of the capacity parameter for the problem being solved is critical. 

\section{Conclusion}
\label{sec:conclusion}

This work addresses the problem of verifiable, privacy-preserving and collaborative MOO on blockchain where mutually distrusting solvers contribute their solutions. In this regard, a novel zk-NSGA-II protocol is proposed, which executes NSGA-II inside a zkVM and certifies each evolutionary generation with a succinct proof that is later verified on-chain. In the protocol, the solvers privately maintain their own populations with decision variables and off-chain Pareto fronts, which are aggregated into a single global Pareto front on-chain. Population states are linked through hash chains, which guarantees that transitions from the current populations to the next populations are carried out by NSGA-II. The correctness of the protocol is first assessed on the ZDT and DTLZ benchmarks with known Pareto fronts. Further, the feasibility of the protocol is shown on the real-world problem of the Hang Seng portfolio optimization. The performance is measured from three complementary perspectives of optimization, ZKP and blockchain with nine different metrics. The experimental measurements indicate that the privacy-preserving collaborative optimization based on NSGA-II is feasible with the existing zero-knowledge and blockchain technologies. However, the cost of proof generation is clearly visible as the primary bottleneck for further adoption.

\bibliographystyle{unsrt}
\bibliography{reference}

\onecolumn
\appendix

\clearpage

\section{zk-NSGA-II Pseudocode}
\label{app:pseudocode}

This appendix gives the pseudocode of the three components in zk-NSGA-II. Algorithm \ref{alg:solver} describes the off-chain loop of a solver, Algorithm \ref{alg:guest} the NSGA-II generation proven in the zkVM and Algorithm \ref{alg:submit} the on-chain verification and aggregation on the smart contract.

\begin{algorithm}[H]
\caption{\textbf{(Host)} Off-chain loop of a solver $i$ with address $a_i$. The step numbers in the comments refer to Figure \ref{fig:architecture}.}
\label{alg:solver}
\small
\begin{algorithmic}[1]
\Require Address $a_i$, guest program $\Gamma$, private population store $\mathcal{D}$, number of generations $G$
\If{$\mathcal{D}$ is empty} \Comment{genesis}
  \State choose an arbitrary population $P_0$ of $N$ solutions in $\Omega$ and a seed $s_0$
  \State $\mathcal{D} \gets \textsc{Store}(P_0, s_0)$
\EndIf
\For{$t = t_0, \dots, G-1$}
  \State $(P_t, s_t) \gets \textsc{Load}(\mathcal{D}, t)$ \Comment{step 2}
  \State $(\pi_t, o_t, P_{t+1}, s_{t+1}) \gets \textsc{Prove}\bigl(\Gamma; \textsc{Guest}(P_t, s_t, a_i)\bigr)$ \Comment{steps 3--6}
  \State $\mathcal{D} \gets \textsc{Store}(P_{t+1}, s_{t+1})$ \Comment{step 7}
  \State send transaction $\textsc{Submit}(\pi_t, o_t)$ from $a_i$ \Comment{step 8}
\EndFor
\end{algorithmic}
\end{algorithm}

\begin{algorithm}[H]
\caption{\textbf{(Guest)} An NSGA-II generation proven in the zkVM as the guest program $\Gamma$.}
\label{alg:guest}
\small
\begin{algorithmic}[1]
\Procedure{Guest}{$P_t$, $a_i$} \Comment{step 3}
  \State $h_t \gets H(P_t)$ \Comment{previous state hash}
  \State $Q_t \gets \varnothing$
  \While{$|Q_t| < N$} \Comment{offspring generation}
    \State $\mathbf{p}_1, \mathbf{p}_2 \gets$ binary tournaments on $P_t$
    \State $\mathbf{c}_1, \mathbf{c}_2 \gets \textsc{SBX}(\mathbf{p}_1, \mathbf{p}_2; p_c, \eta_c)$ \Comment{simulated binary crossover}
    \State $\mathbf{c}_1, \mathbf{c}_2 \gets \textsc{PM}(\mathbf{c}_1, \mathbf{c}_2; p_m, \eta_m)$ \Comment{polynomial mutation}
    \State $Q_t \gets Q_t \cup \{\mathbf{c}_1, \mathbf{c}_2\}$
  \EndWhile
  \State $\mathcal{F}_1, \mathcal{F}_2, \dots \gets \textsc{NonDominatedSort}(P_t \cup Q_t)$ \Comment{sort merged population}
  \State $P_{t+1} \gets \varnothing$; $j \gets 1$
  \While{$|P_{t+1}| + |\mathcal{F}_j| \leq N$} \Comment{front by front survivor selection}
    \State $P_{t+1} \gets P_{t+1} \cup \mathcal{F}_j$; $j \gets j + 1$
  \EndWhile
  \State $h_{t+1} \gets H(P_{t+1})$ \Comment{new state hash}
  \State $\mathbf{x}^{\ast}_{t+1} \gets$ solution of $\mathcal{F}_1(P_{t+1})$ at index randomly drawn 
  \State \textbf{commit} $o_t = (a_i, h_t, h_{t+1}, \mathbf{F}(\mathbf{x}^{\ast}_{t+1}))$ \Comment{step 6}
  \State \Return $P_{t+1}$ \Comment{step 5}
\EndProcedure
\end{algorithmic}
\end{algorithm}

\begin{algorithm}[H]
\caption{\textbf{(Smart Contract)} Front aggregation by the Pareto Front contract. The step numbers in the comments refer to Figure \ref{fig:architecture}.}
\label{alg:submit}
\small
\begin{algorithmic}[1]
\Procedure{Submit}{$\pi_t$, $o_t$} \Comment{step 8}
  \State \textbf{require} $\textsc{Verify}(\Gamma; \pi_t, o_t)$ \Comment{step 9}
  \State \textbf{require} $a_i = \texttt{msg.sender}$ \Comment{solver check}
  \State \textbf{require} $\mathrm{head}[a_i] = \bot$ \textbf{or} $h_t = \mathrm{head}[a_i]$ \Comment{hash-chain check}
  \State $\mathrm{head}[a_i] \gets h_{t+1}$ \Comment{advance the hash chain}
  \If{no $\mathbf{q} \in \mathcal{G}$ dominates or equals $\mathbf{F}(\mathbf{x}^{\ast}_{t+1})$}
    \State remove from $\mathcal{G}$ all points dominated by $\mathbf{F}(\mathbf{x}^{\ast}_{t+1})$
    \If{$|\mathcal{G}| < K$ \textbf{or} a point is removed} \Comment{capacity check}
      \State $\mathcal{G} \gets \mathcal{G} \cup \{\mathbf{F}(\mathbf{x}^{\ast}_{t+1})\}$
      \State \textbf{emit} \texttt{admitted}$(a_i, \mathbf{F}(\mathbf{x}^{\ast}_{t+1}))$
      \State \Return
    \EndIf
  \EndIf
  \State \textbf{emit} \texttt{discarded}$(a_i, \mathbf{F}(\mathbf{x}^{\ast}_{t+1}))$ \Comment{dominated or front full}
\EndProcedure
\end{algorithmic}
\end{algorithm}

\clearpage
\section{Definitions of the Benchmark Problems}
\label{app:problems}

Tables \ref{tab:zdt} and \ref{tab:dtlz} give the definitions of the ZDT and DTLZ problems used in Section \ref{sec:problems}.

\begin{table}[H]
\caption{Definitions of the ZDT problems \cite{zitzler2000} used in this work. Both objectives are minimized and $n$ denotes the number of decision variables. The Pareto front is given in the last column.}
\label{tab:zdt}
\small
\begin{tabularx}{\textwidth}{@{}l c l X p{3.6cm}@{}}
\toprule
\textbf{Problem} & $n$ & \textbf{Bounds} & \textbf{Objective functions} & \textbf{Pareto front} \\
\midrule
\addlinespace[2pt]
ZDT1 & 30 & $x_j \in [0, 1]$ &
$f_1 = x_1$ \newline $f_2 = g \left(1 - \sqrt{f_1 / g}\right)$ \newline $g = 1 + \frac{9}{n-1} \sum_{j=2}^{n} x_j$ &
$f_2 = 1 - \sqrt{f_1}$ \newline convex \\
\addlinespace[4pt]
ZDT2 & 30 & $x_j \in [0, 1]$ &
$f_1 = x_1$ \newline $f_2 = g \left(1 - (f_1 / g)^2\right)$ \newline $g = 1 + \frac{9}{n-1} \sum_{j=2}^{n} x_j$ &
$f_2 = 1 - f_1^2$ \newline non-convex \\
\addlinespace[4pt]
ZDT3 & 30 & $x_j \in [0, 1]$ &
$f_1 = x_1$ \newline $f_2 = g \left(1 - \sqrt{f_1 / g} - \frac{f_1}{g} \sin(10 \pi f_1)\right)$ \newline $g = 1 + \frac{9}{n-1} \sum_{j=2}^{n} x_j$ &
$f_2 = 1 - \sqrt{f_1} - f_1 \sin(10 \pi f_1)$ \newline disconnected \\
\addlinespace[4pt]
ZDT4 & 10 & \begin{tabular}[t]{@{}l@{}}$x_1 \in [0, 1]$ \\ $x_j \in [-5, 5]$, $j \geq 2$\end{tabular} &
$f_1 = x_1$ \newline $f_2 = g \left(1 - \sqrt{f_1 / g}\right)$ \newline $g = 1 + 10(n-1) + \sum_{j=2}^{n} \left(x_j^2 - 10 \cos(4 \pi x_j)\right)$ &
$f_2 = 1 - \sqrt{f_1}$ \newline convex, multi-modal \\
\addlinespace[4pt]
ZDT6 & 10 & $x_j \in [0, 1]$ &
$f_1 = 1 - e^{-4 x_1} \sin^{6}(6 \pi x_1)$ \newline $f_2 = g \left(1 - (f_1 / g)^2\right)$ \newline $g = 1 + 9 \left(\frac{1}{n-1} \sum_{j=2}^{n} x_j\right)^{0.25}$ &
$f_2 = 1 - f_1^2$ \newline non-convex, non-uniform \\
\bottomrule
\end{tabularx}
\end{table}

\begin{table}[H]
\caption{Definitions of the DTLZ problems \cite{deb2002} used in this work. Both objectives are minimized, $n$ denotes the number of decision variables and $k = n - 1$. All decision variables lie in $[0, 1]$. The Pareto front is given in the last column.}
\label{tab:dtlz}
\small
\begin{tabularx}{\textwidth}{@{}l c X p{3.6cm}@{}}
\toprule
\textbf{Problem} & $n$ & \textbf{Objective functions} & \textbf{Pareto front} \\
\midrule
\addlinespace[2pt]
DTLZ1 & 6 &
$f_1 = \frac{1}{2} x_1 (1 + g)$ \newline $f_2 = \frac{1}{2} (1 - x_1)(1 + g)$ \newline $g = 100 \left[ k + \sum_{j=2}^{n} \left( (x_j - 0.5)^2 - \cos\left(20 \pi (x_j - 0.5)\right) \right) \right]$ &
$f_1 + f_2 = \frac{1}{2}$ \newline linear, multi-modal \\
\addlinespace[4pt]
DTLZ2 & 11 &
$f_1 = (1 + g) \cos\left(\frac{\pi}{2} x_1\right)$ \newline $f_2 = (1 + g) \sin\left(\frac{\pi}{2} x_1\right)$ \newline $g = \sum_{j=2}^{n} (x_j - 0.5)^2$ &
$f_1^2 + f_2^2 = 1$ \newline non-convex \\
\addlinespace[4pt]
DTLZ3 & 11 &
$f_1 = (1 + g) \cos\left(\frac{\pi}{2} x_1\right)$ \newline $f_2 = (1 + g) \sin\left(\frac{\pi}{2} x_1\right)$ \newline $g = 100 \left[ k + \sum_{j=2}^{n} \left( (x_j - 0.5)^2 - \cos\left(20 \pi (x_j - 0.5)\right) \right) \right]$ &
$f_1^2 + f_2^2 = 1$ \newline non-convex, multi-modal \\
\addlinespace[4pt]
DTLZ4 & 11 &
$f_1 = (1 + g) \cos\left(\frac{\pi}{2} x_1^{\alpha}\right)$ \newline $f_2 = (1 + g) \sin\left(\frac{\pi}{2} x_1^{\alpha}\right)$, $\alpha = 100$ \newline $g = \sum_{j=2}^{n} (x_j - 0.5)^2$ &
$f_1^2 + f_2^2 = 1$ \newline non-convex, biased density \\
\addlinespace[4pt]
DTLZ5 & 11 &
$f_1 = (1 + g) \cos\left(\frac{\pi}{2} x_1\right)$ \newline $f_2 = (1 + g) \sin\left(\frac{\pi}{2} x_1\right)$ \newline $g = \sum_{j=2}^{n} (x_j - 0.5)^2$ &
$f_1^2 + f_2^2 = 1$ \newline non-convex, identical to DTLZ2 \\
\addlinespace[4pt]
DTLZ6 & 11 &
$f_1 = (1 + g) \cos\left(\frac{\pi}{2} x_1\right)$ \newline $f_2 = (1 + g) \sin\left(\frac{\pi}{2} x_1\right)$ \newline $g = \sum_{j=2}^{n} x_j^{0.1}$ &
$f_1^2 + f_2^2 = 1$ \newline non-convex, biased distance \\
\addlinespace[4pt]
DTLZ7 & 21 &
$f_1 = x_1$ \newline $f_2 = (1 + g) \left( 2 - \frac{f_1}{1 + g} \left(1 + \sin(3 \pi f_1)\right) \right)$ \newline $g = 1 + \frac{9}{k} \sum_{j=2}^{n} x_j$ &
$f_2 = 4 - f_1 \left(1 + \sin(3 \pi f_1)\right)$ \newline disconnected \\
\bottomrule
\end{tabularx}
\end{table}

\clearpage
\section{Acceptance Rate and Global Front Size}
\label{app:acceptance}

\begin{figure}[H]
\centering
\begin{subfigure}[t]{0.245\textwidth}\centering\includegraphics[width=\linewidth]{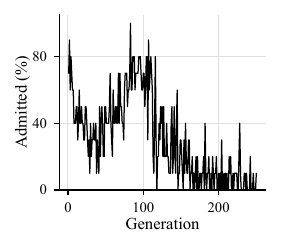}\caption{ZDT1}\label{fig:res-acceptance-zdt1}\end{subfigure}
\hfill
\begin{subfigure}[t]{0.245\textwidth}\centering\includegraphics[width=\linewidth]{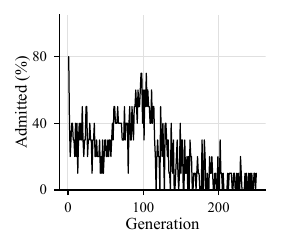}\caption{ZDT2}\label{fig:res-acceptance-zdt2}\end{subfigure}
\hfill
\begin{subfigure}[t]{0.245\textwidth}\centering\includegraphics[width=\linewidth]{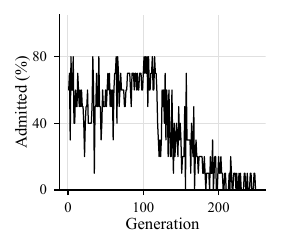}\caption{ZDT3}\label{fig:res-acceptance-zdt3}\end{subfigure}
\hfill
\begin{subfigure}[t]{0.245\textwidth}\centering\includegraphics[width=\linewidth]{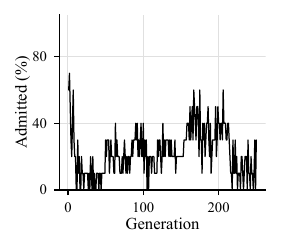}\caption{ZDT4}\label{fig:res-acceptance-zdt4}\end{subfigure}
\\[2pt]
\begin{subfigure}[t]{0.245\textwidth}\centering\includegraphics[width=\linewidth]{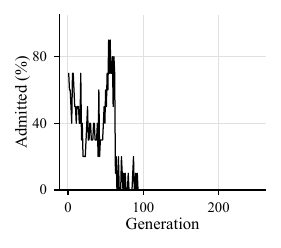}\caption{ZDT6}\label{fig:res-acceptance-zdt6}\end{subfigure}
\hfill
\begin{subfigure}[t]{0.245\textwidth}\centering\includegraphics[width=\linewidth]{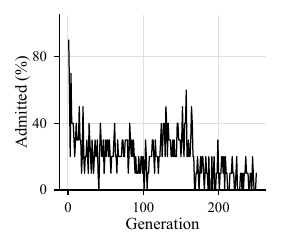}\caption{DTLZ1}\label{fig:res-acceptance-dtlz1}\end{subfigure}
\hfill
\begin{subfigure}[t]{0.245\textwidth}\centering\includegraphics[width=\linewidth]{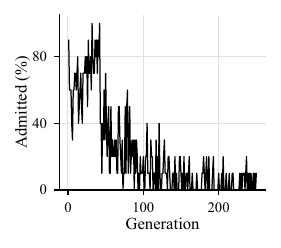}\caption{DTLZ2}\label{fig:res-acceptance-dtlz2}\end{subfigure}
\hfill
\begin{subfigure}[t]{0.245\textwidth}\centering\includegraphics[width=\linewidth]{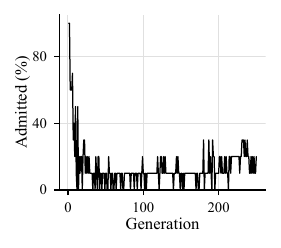}\caption{DTLZ3}\label{fig:res-acceptance-dtlz3}\end{subfigure}
\\[2pt]
\begin{subfigure}[t]{0.245\textwidth}\centering\includegraphics[width=\linewidth]{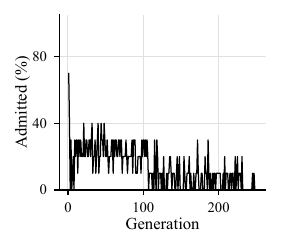}\caption{DTLZ4}\label{fig:res-acceptance-dtlz4}\end{subfigure}
\hfill
\begin{subfigure}[t]{0.245\textwidth}\centering\includegraphics[width=\linewidth]{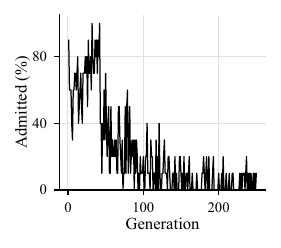}\caption{DTLZ5}\label{fig:res-acceptance-dtlz5}\end{subfigure}
\hfill
\begin{subfigure}[t]{0.245\textwidth}\centering\includegraphics[width=\linewidth]{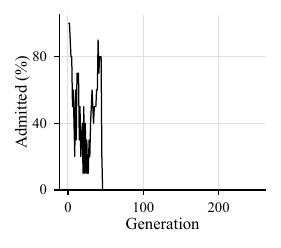}\caption{DTLZ6}\label{fig:res-acceptance-dtlz6}\end{subfigure}
\hfill
\begin{subfigure}[t]{0.245\textwidth}\centering\includegraphics[width=\linewidth]{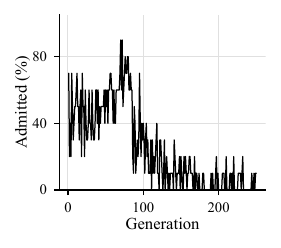}\caption{DTLZ7}\label{fig:res-acceptance-dtlz7}\end{subfigure}
\\[2pt]
\caption{Acceptance rate of the global front over the $G = 250$ generations for the twelve ZDT and DTLZ problems. At each generation, each solver submits one objective vector for the admission to the global front.}
\label{fig:res-acceptance}
\end{figure}

\begin{figure}[H]
\centering
\begin{subfigure}[t]{0.245\textwidth}\centering\includegraphics[width=\linewidth]{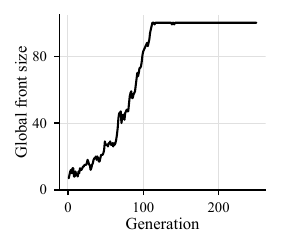}\caption{ZDT1}\label{fig:res-size-zdt1}\end{subfigure}
\hfill
\begin{subfigure}[t]{0.245\textwidth}\centering\includegraphics[width=\linewidth]{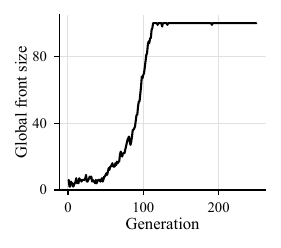}\caption{ZDT2}\label{fig:res-size-zdt2}\end{subfigure}
\hfill
\begin{subfigure}[t]{0.245\textwidth}\centering\includegraphics[width=\linewidth]{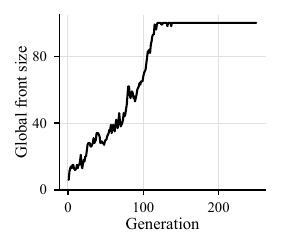}\caption{ZDT3}\label{fig:res-size-zdt3}\end{subfigure}
\hfill
\begin{subfigure}[t]{0.245\textwidth}\centering\includegraphics[width=\linewidth]{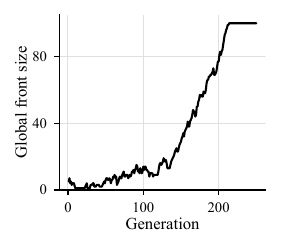}\caption{ZDT4}\label{fig:res-size-zdt4}\end{subfigure}
\\[2pt]
\begin{subfigure}[t]{0.245\textwidth}\centering\includegraphics[width=\linewidth]{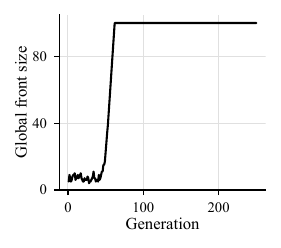}\caption{ZDT6}\label{fig:res-size-zdt6}\end{subfigure}
\hfill
\begin{subfigure}[t]{0.245\textwidth}\centering\includegraphics[width=\linewidth]{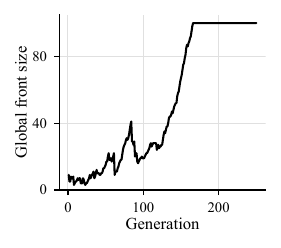}\caption{DTLZ1}\label{fig:res-size-dtlz1}\end{subfigure}
\hfill
\begin{subfigure}[t]{0.245\textwidth}\centering\includegraphics[width=\linewidth]{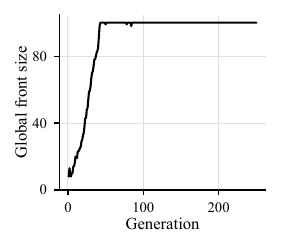}\caption{DTLZ2}\label{fig:res-size-dtlz2}\end{subfigure}
\hfill
\begin{subfigure}[t]{0.245\textwidth}\centering\includegraphics[width=\linewidth]{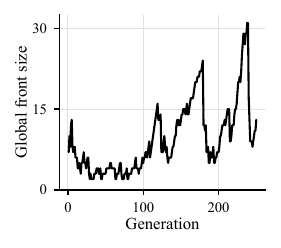}\caption{DTLZ3}\label{fig:res-size-dtlz3}\end{subfigure}
\\[2pt]
\begin{subfigure}[t]{0.245\textwidth}\centering\includegraphics[width=\linewidth]{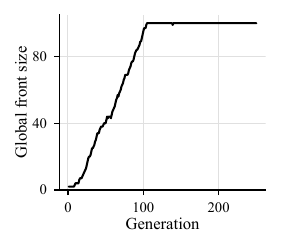}\caption{DTLZ4}\label{fig:res-size-dtlz4}\end{subfigure}
\hfill
\begin{subfigure}[t]{0.245\textwidth}\centering\includegraphics[width=\linewidth]{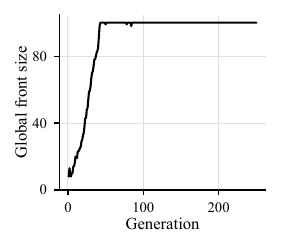}\caption{DTLZ5}\label{fig:res-size-dtlz5}\end{subfigure}
\hfill
\begin{subfigure}[t]{0.245\textwidth}\centering\includegraphics[width=\linewidth]{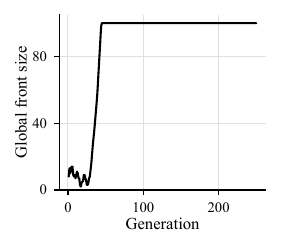}\caption{DTLZ6}\label{fig:res-size-dtlz6}\end{subfigure}
\hfill
\begin{subfigure}[t]{0.245\textwidth}\centering\includegraphics[width=\linewidth]{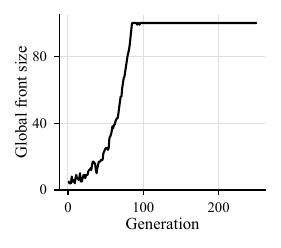}\caption{DTLZ7}\label{fig:res-size-dtlz7}\end{subfigure}
\\[2pt]
\caption{Size of the on-chain global front over the $G = 250$ generations for the twelve ZDT and DTLZ problems with a capacity of $K = 100$. The size grows as solutions are admitted after proof verification and drops when a new solution removes the ones it dominates. Once the size reaches the capacity $K$, a new solution is admitted if it replaces at least one existing solution.}
\label{fig:res-size}
\end{figure}

Figures \ref{fig:res-acceptance} and \ref{fig:res-size} complement the aggregation results in Section \ref{sec:res-local-global} by showing how the acceptance rate and the global front size change over the generations of the ZDT and DTLZ problems.

\end{document}